\documentclass[
reprint,
amsmath,amssymb,
aps,prb,
superscriptaddress,
floatfix
]{revtex4-1}

\usepackage{graphicx}
\usepackage{braket}

\begin{document}

\title{Nature of Protected Zero Energy States in Penrose Quasicrystals}  
\author{Ezra Day-Roberts}
\affiliation{School of Physics and Astronomy, University of Minnesota, Minneapolis, MN 55455, USA}
\author{Rafael M. Fernandes}
\affiliation{School of Physics and Astronomy, University of Minnesota, Minneapolis, MN 55455, USA}
\author{Alex Kamenev}
\affiliation{School of Physics and Astronomy, University of Minnesota, Minneapolis, MN 55455, USA}
\affiliation{William I. Fine Theoretical Physics Institute, University of Minnesota, Minneapolis, Minnesota 55455, USA}
\begin{abstract}
The electronic spectrum of the Penrose rhombus quasicrystal exhibits a macroscopic fraction of exactly degenerate zero energy states. In contrast to other bipartite quasicrystals, such as the kite-and-dart one, these zero energy states cannot be attributed to a global mismatch $\Delta n$ between the number of sites in the two sublattices that form the quasicrystal. Here, we argue that these zero energy states are instead related to a \emph{local} mismatch $\Delta n(\bf r)$. Although $\Delta n(\bf r)$ averages to zero, its staggered average over self-organized domains gives the correct number of zero energy states. Physically, the local mismatch is related to a hidden structure of nested self-similar domains that support the zero energy states. This allows us to develop a real space renormalization-group scheme, which yields the scaling law for the fraction of zero energy states, $Z$, versus size of their support domain, $N$, as $Z\propto N^{-\eta}$ with $\eta =1-\ln 2/\ln(1+\tau) \approx 0.2798$ (where $\tau$ is the golden ratio). It also reproduces the known total fraction of the zero energy states, $81-50\tau\approx 0.0983$.  We also show that the exact degeneracy of these states is protected against a wide variety of local perturbations, such as irregular or random hopping amplitudes, magnetic field, random dilution of the lattice, etc. We attribute this robustness to the hidden domain structure and speculate about its underlying topological origin. 
\end{abstract}
\maketitle

\section{Introduction}
Quasicrystals were first discovered in 1984 by Shechtman et al. \cite{original} in Al alloys. Within a short time, many other quasiperiodic crystals were discovered\cite{PhysRevLett.55.511,PhysRevLett.59.1010,PhysRevLett.55.1461} including, eventually, a naturally occurring AlCuFe quasicrystal  \cite{Bindi1306}. More recent work has focused on connecting quasicrystals with other novel phenomena such as topological states \cite{topological-states,PhysRevLett.122.095501,top-states-2,PhysRevLett.119.215304}, non-Fermi liquid behavior  \cite{Watanuki12,PhysRevB.96.241102,heavy-fermion,JPSJ.85.063706}, superconductivity \cite{PhysRevB.95.024509,Araujo19,Zhang2020,Cao2020}, and quantum criticality \cite{Deguchi2012,Scalettar16}. Synthetic quasicrystals were also recently created by arranging CO molecules on a Cu$(111)$ surface with the aid of scanning tunneling microscopy \cite{Collins2017}. 

Quasicrystals are known to display unusual properties in their density of states. All 1D quasicrystals have a density of states (DOS) that is only non-zero on a set of measure zero\cite{Lenz-zero-weight,Damanik1999}. The simplest 2D quasicrystals, which are built as cartesian products of 1D quasicrystals, have densities of states that are related to those of their 1D counterparts \cite{square-and-cubic-fib,octagonal}. On the other hand, the DOS of intrinsic 2D and 3D quasicrystals can be rather different, displaying a sharp suppression at the Fermi level or a macroscopic number of zero energy states \cite{PhysRevLett.66.333,PhysRevLett.79.1070,PhysRevB.68.014205}.

In this paper, we focus on the nearest-neighbor tight-binding model on the Penrose rhombus lattice, which is known to display a macroscopic number of zero energy states. Several properties of these zero energy states are well-understood, including their fraction in the thermodynamic limit, $f=81-50\tau \approx 9.8\%$, where $\tau = (\sqrt{5}+1)/2$ is the golden ratio \cite{Kohmoto_Sutherland,nine_percent,exact-number}. However, other properties are not as clear, such as their microscopic origin and their stability against perturbations. Indeed, in nearest-neighbor tight-binding models on bipartite lattices, such as the Penrose lattice, zero energy states can be trivially generated if the number of sites on the two sublattices, which we will refer to as $A$ and $B$, are not the same. In other words, there is a global mismatch $\Delta n \equiv n_A - n_B \neq 0$, implying the existence of a number of zero energy states equal to $|\Delta n|$. This is precisely the case in the bipartite quasicrystal known as the Penrose kite-and-dart lattice, whose sublattice site mismatch gives rise to $\approx 10\%$ of zero energy states. This is of course by no means limited to quasicrystals; for instance, the dice lattice has one third of its eigenstates at zero energy, reflecting the sublattice mismatch in each unit cell.

\begin{figure}
\centering
\includegraphics[width=0.4\textwidth]{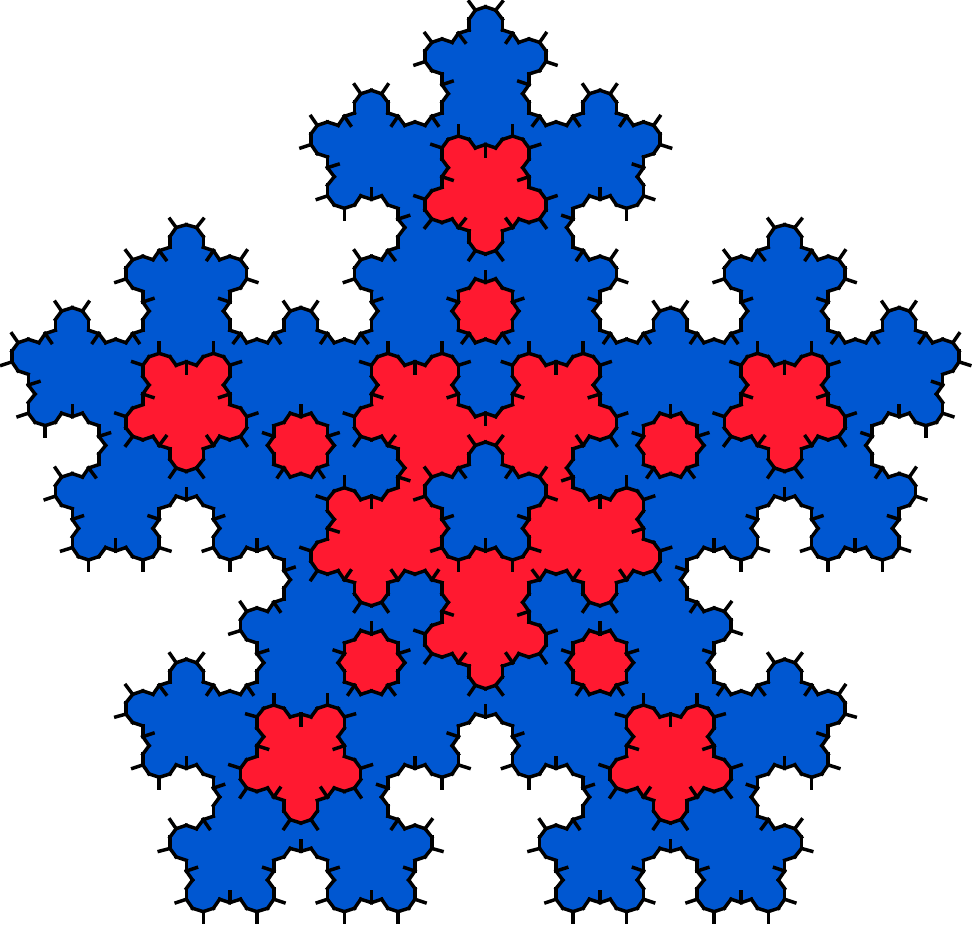}
\caption{A section of approximately $4,500$ sites of the Penrose rhombus lattice divided into domains showing the nesting of sub-domains. In the red (blue) domains, the local sublattice mismatch is such that the A (B) sublattice has more sites than the B (A) sublattice. The domain walls connect sites belonging to different sublattices.}
\label{fig:nested-domains}
\end{figure}

In the case of the Penrose rhombus lattice, however, there is no global mismatch in the thermodynamic limit, $\Delta n = 0$. This raises an important question about the nature of the zero energy states. To address this issue, in this paper we introduce the concept of a \emph{local} sublattice mismatch, $\Delta n(\bf r)$. The key point is that, even though the average $\langle \Delta n(\bf r) \rangle$ is zero, the staggered average $\langle (-1)^{\mathcal{S}_{\bf r}} \Delta n(\bf r) \rangle$, in which $\Delta n(\bf r)$ changes sign across certain regions denoted by $\mathcal{S}_{\bf r}$, can be non-zero. The situation is analogous to an antiferromagnet: while the average magnetization vanishes, the average staggered magnetization is finite. 

However, in contrast to an antiferromagnet, where the magnetization changes sign at the atomic length scale, the local mismatch $\Delta n(\bf r)$ changes sign at much larger scales in the Penrose lattice. This is illustrated in Fig.~\ref{fig:nested-domains}, which shows the spatial variation of $\Delta n(\bf r)$ for a Penrose lattice with about 4,500 sites. In the red (blue) ``domains", the local sublattice mismatch is such that $n_A > n_B$ ($n_A < n_B$). Defining a staggered $\Delta n(\bf r)$ by changing its sign in the blue domains as compared to the red domains yields a finite number that coincides with the number of zero energy states. Interestingly, a recent investigation of the classical dimer model on the Penrose lattice found a result that resembles ours, namely, that the Penrose lattice supports a cluster structure with charge-alternating monomers, despite the fact that the Penrose lattice itself is globally charge neutral \cite{dimers}. 

The connection between the staggered $\Delta n(\bf r)$ and the zero energy states can be made more transparent by considering the excluded sites (or forbidden sites, as they were originally called in Ref. \onlinecite{nine_percent}), i.e. sites for which the zero energy states wave-functions vanish. While inside the red domains the excluded sites are all in the A sublattice, in the blue domains they switch to the B sublattice. The domain walls (called strings in Ref. \onlinecite{nine_percent}) therefore connect excluded sites that belong to opposite sublattices.

We emphasize that this hidden geometric structure of the zero energy states was already noted in several previous works \cite{nine_percent,HATAKEYAMA198779,exact-number,PhysRevB.95.024509,dimers}. One of our main points here is to connect this structure to a staggered local mismatch $\Delta n(\bf r)$ that spontaneously forms in the Penrose lattice. This geometric structure protects the zero energy states from any perturbation that does not disrupt the large scale nearest-neighbor structure. This includes, as was previously found, a perpendicular magnetic field\cite{HATAKEYAMA198779}. More generally, we demonstrate robustness against random nearest-neighbor hopping amplitudes and single-site vacancies. Even the addition of further hopping terms that break the bipartite symmetry, like next-nearest-neighbors, only reduces the number of zero energy states linearly in the number of next-nearest-neighbor hopping amplitudes.

The domain structure of the local mismatch $\Delta n(\bf r)$, combined with the inflation properties of the Penrose lattice, also allow us to derive a recursive equation for the increase of the staggered mismatch as a function of the Penrose lattice generation. The structure of such a recursive relation resembles a real-space renormalization-group (RG) flow. Previous works have applied the real-space RG technique to solve the Ising model on the Penrose lattice and to compute the local density of states  \cite{aoyama1987eight,You_1992}; here, however, our goal is to determine the staggered local mismatch. We solve the flow equations to find the staggered mismatch in the infinite lattice limit. We find that the total number of staggered mismatched sites corresponds to a fraction $f=81-50\tau \approx 0.0983$ of sites \cite{exact-number}, showing that all zero energy states in the Penrose lattice originate from this local mismatch structure.  Using the real-space RG, we also find a scaling law $Z\propto N^{-\eta}$ that relates the fraction of zero energy states, $Z$, to the size of their support domain, $N$. Our calculations give an exponent $\eta =1-\ln 2/\ln(1+\tau) \approx 0.2798$.

The paper is organized as follows. Section \ref{sec:phenomenology} contrasts the nature of the zero energy states and of the global sublattice mismatch in two different Penrose quasicrystals, the kite-and-dart lattice and the rhombus lattice, and introduces the domain structure of the local mismatch in the rhombus lattice.
Section \ref{sec:robustness} discusses robustness of the zero energy states against various perturbations as well as the spatial structure of the zero energy states. Section \ref{sec:RG} uses the inflation property of the Penrose lattice to derive a real-space RG-like approach to determine the number of zero energy states by exploiting its connection to the number of locally mismatched sublattice sites. Section \ref{sec:conclusions} concludes the paper by discussing possible topological aspects of the zero energy states. Appendices A, B, C, and D contain details about the derivation of the RG-like recursive relations.

\section{Phenomenology of zero energy states in quasicrystals} \label{sec:phenomenology}

We consider tight-binding Hamiltonians defined on 2D quasicrystal lattices with zero on-site energies and nearest-neighbor hopping. Such Hamiltonians are numerically diagonalized on lattices containing up to $11,000$ sites. 
Below we summarize our findings for the kite-and-dart and rhombus lattices. It is important to stress that both these lattices  are {\em bipartite} with all sites belonging to either A or B sublattice and nearest neighbor hoping operating solely between them.   

\subsection{The Kite-and-Dart Lattice}
\begin{figure}
\centering
\includegraphics[width=.3\textwidth]{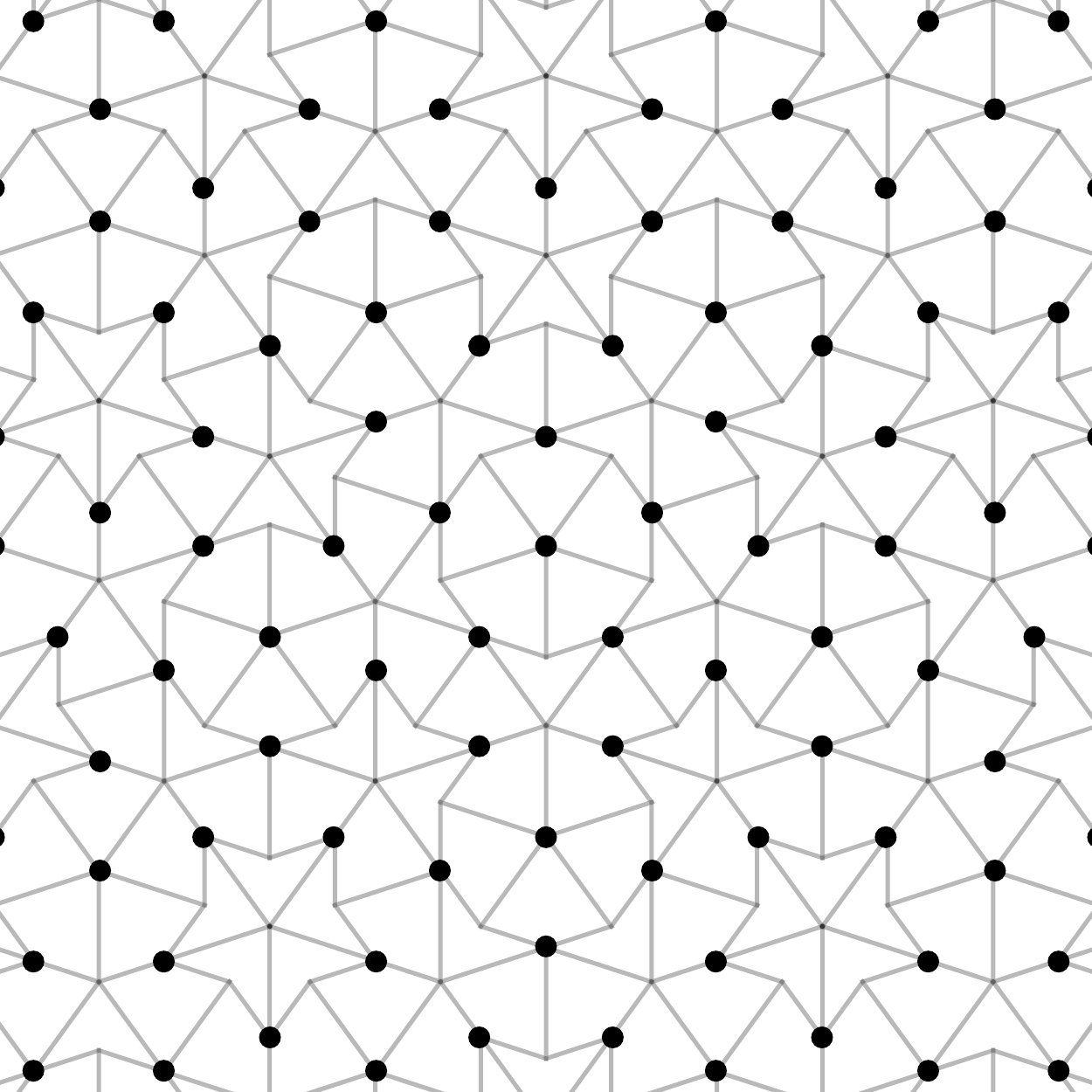}
\caption{Kite-and-dart lattice. Sites with {\em no} amplitude of {\em any} of the zero energy states are marked by the black dots. Notice that  all marked sites belong to one sublattice.}
\label{fig:kd}
\end{figure}

\begin{figure}
\includegraphics[width=.45\textwidth]{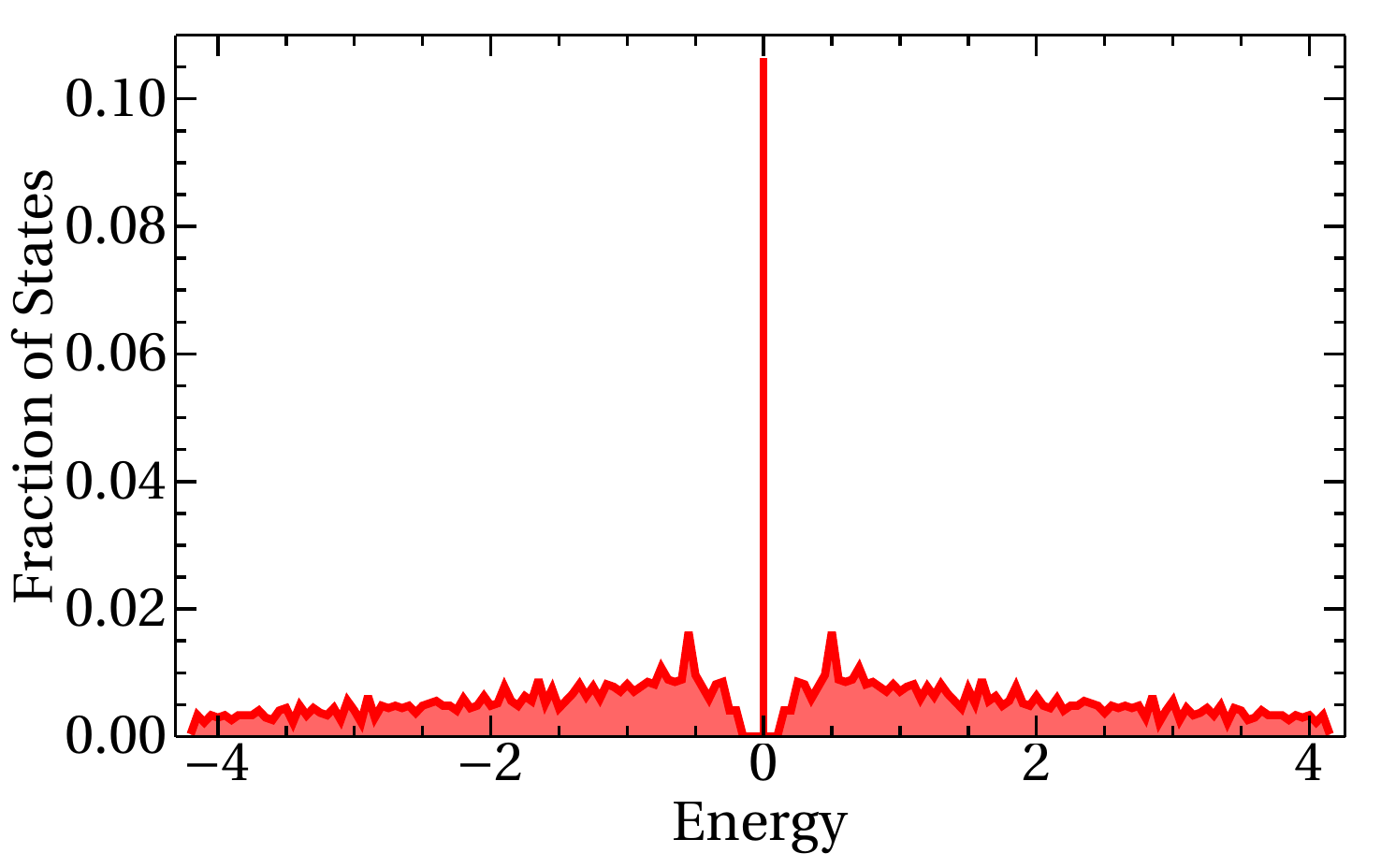}
\caption{Density of states for the kite-and-dart lattice. DOS is symmetric about zero and has a macroscopic number ($\sim10\%$) of exactly zero energy states.}
\label{fig:kd-DOS}
\end{figure}

Figure \ref{fig:kd} shows a section of the kite-and-dart lattice. It is a quasicrystal with a five-fold rotational symmetry. The corresponding density of states (DOS)  is shown in Fig.~\ref{fig:kd-DOS}, where a large peak of zero energy states is clearly visible. 

\begin{table}
\begin{ruledtabular}
\begin{tabular}{l|l|l}
 \textrm{$\#$ of sites} & \textrm{$\#$ of zero energy states} & \textrm{(global) mismatch} \\\hline
 166 & 24 & 24 \\
 411 & 41 & 41 \\
 1046 & 104 & 104 \\
 2686 & 286 & 286 \\
 6951 & 739 & 739 \\
\end{tabular}
\end{ruledtabular}
\caption{\label{t:kd_numbers}Numbers of zero energy states and lattice mismatch in kite-and-dart lattices. The zero energy states is completely explained by the (global) sublattice mismatch.}
\end{table}

For this lattice the number of zero energy states is entirely explained by the mismatch in the number of sites of the two sublattices, as seen in table \ref{t:kd_numbers}. As expected for these zero energy states, the amplitude of their wave-functions is non-zero only on the majority sublattice. If the Hamiltonian is written in the sublattice basis it takes the form of a block off-diagonal matrix:
\begin{equation}
\label{eq:off-diagonal} 
\hat H = \begin{pmatrix}
	0 & G \\ G^T & 0
\end{pmatrix}%
\end{equation}
If the two sublattices have different numbers of sites, $G$ has more columns than rows and so must have a null space at least as large as the difference between sizes of the two sublattices. So there must be vector(s) $\vec x$ such that $G \vec x = 0$. Each of these corresponds to an eigenstate, $\psi_0 = (0, \vec x)^T$ of the Hamiltonian $\hat H$ with the zero eigenvalue, $\hat H \psi_0=0$. Thus the zero energy eigenstates have amplitudes only on the majority sublattice and {\em no} amplitude on the minority sublattice. In Fig.~\ref{fig:kd} the sites that have {\it no} amplitude are marked with black dots, which are indeed 
spanning the minority sublattice. Here the mismatch between the two global sublattices, $\Delta n$, exactly accounts for the $\sim10\%$ of all the states having the zero energy. 

This situation is to some extent similar to e.g. the dice lattice, which is bipartite with three sites per unit cell. Two sites belong to the majority and one to the minority sublattice. The mismatch is the third of all the sites and thus $1/3$ of all the states are at exactly zero energy (the so-called flat band). The difference is, of course, that the quasicrystal is not translationally invariant and the states are not labelled by the quasi momentum.   

\subsection{The Rhombus Lattice} 
\begin{figure}[htp]
\centering
\includegraphics[width=0.4\textwidth]{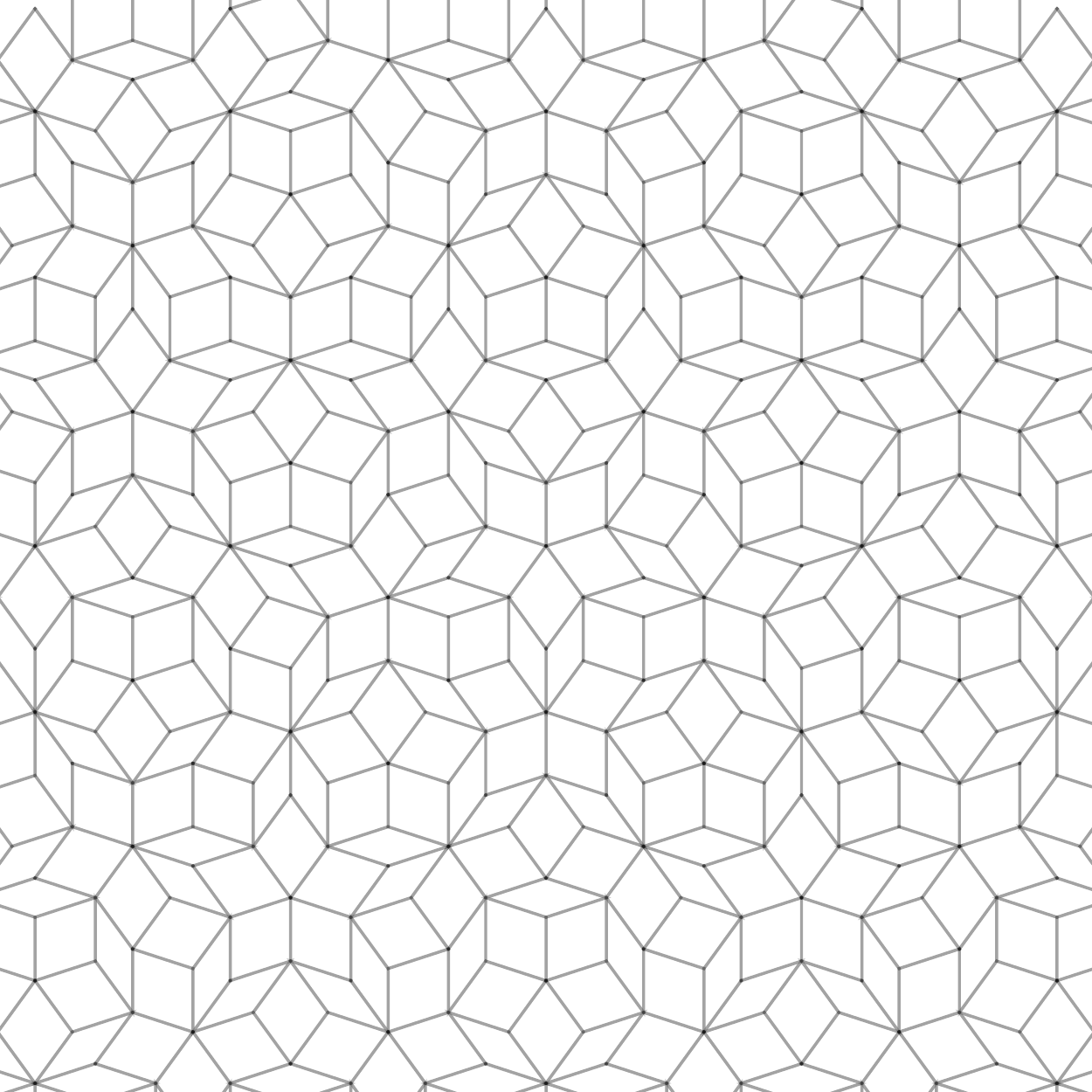}
\caption{The rhombus tiling is bipartite and has five-fold symmetry.}
\label{fig:rhombus}
\end{figure}

\begin{figure}
\centering
\includegraphics[width=0.45\textwidth]{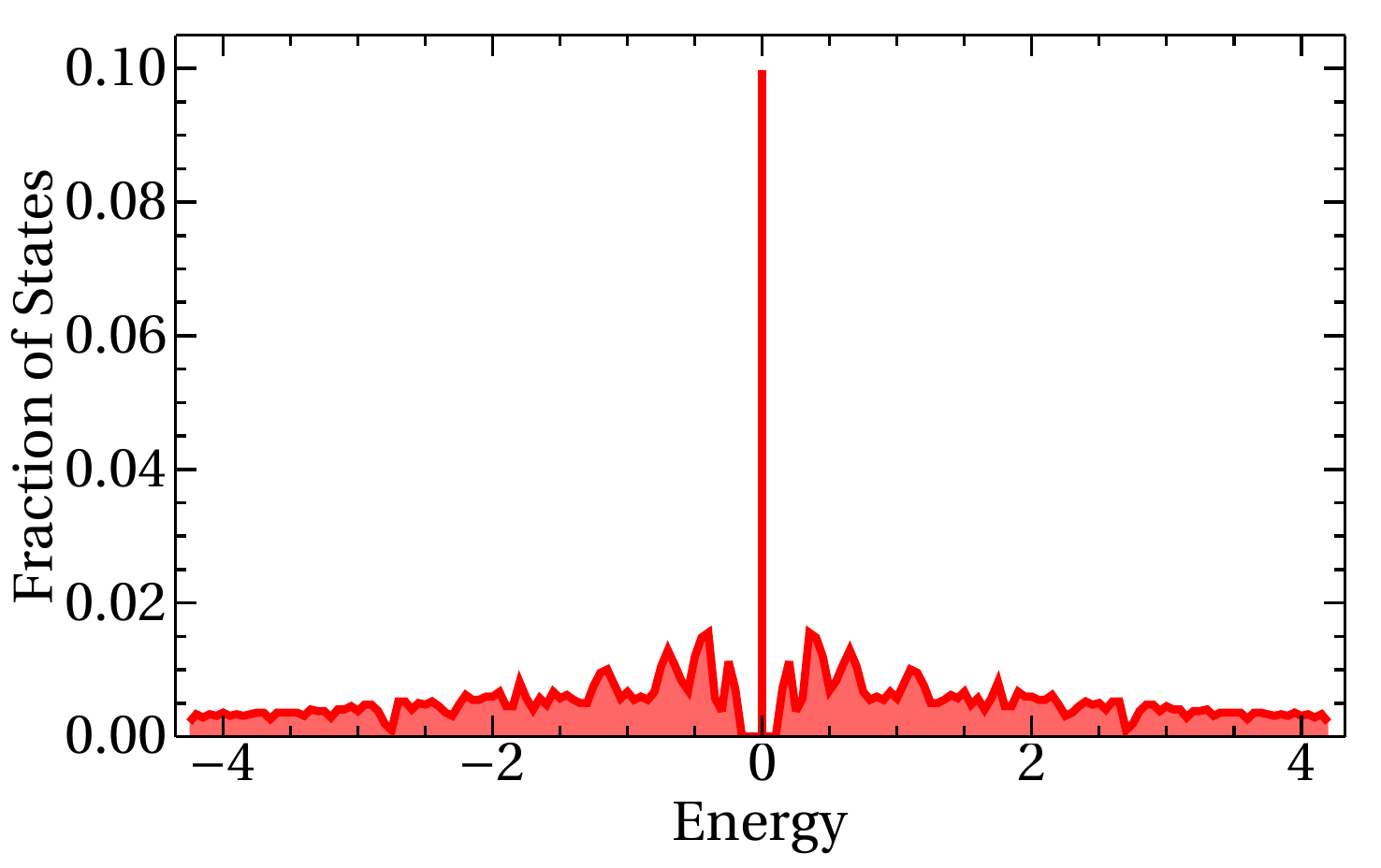}
\caption{The DOS of the rhombus lattice shown here is similar to the DOS of the kite-and-dart lattice in Fig.~\ref{fig:kd-DOS}, however the origin of the zero energy peak is completely different.}
\label{fig:rhomb-DOS}
\end{figure}

\begin{figure}
\centering
\includegraphics[width=0.4\textwidth]{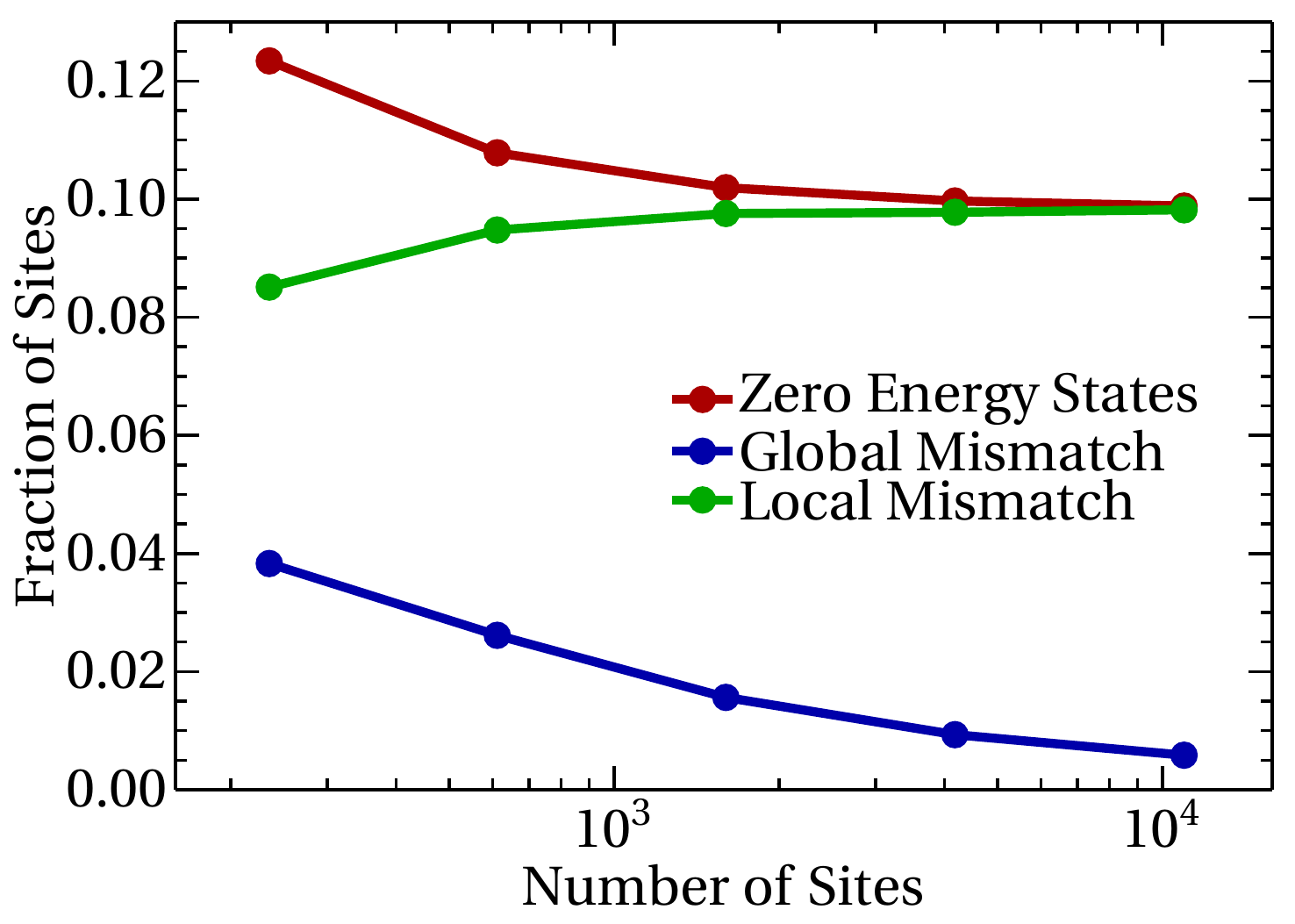}
\caption{Fractions for the mismatch in the number of sites in the two sublattices (global and local, as defined in the main text) and for the number of zero energy states as a function of the total number of sites in the rhombus lattice.}
\label{fig:rhombZESplot}
\end{figure}

The Penrose rhombus lattice is displayed Fig.~\ref{fig:rhombus} and its DOS is shown in Fig.~\ref{fig:rhomb-DOS}. Similarly to the kite-and-dart lattice, there is about $10\%$ of exactly zero energy states. However, in this case this number cannot be explained by the mismatch between the two sublattices.  As seen in Fig.~\ref{fig:rhombZESplot} (see global mismatch line) the global relative sublattice mismatch $\Delta n$ goes to zero, while the fraction of the zero energy states saturates to a constant upon increasing the lattice size. 
 
\begin{figure}
\centering
\includegraphics[width=0.4\textwidth]{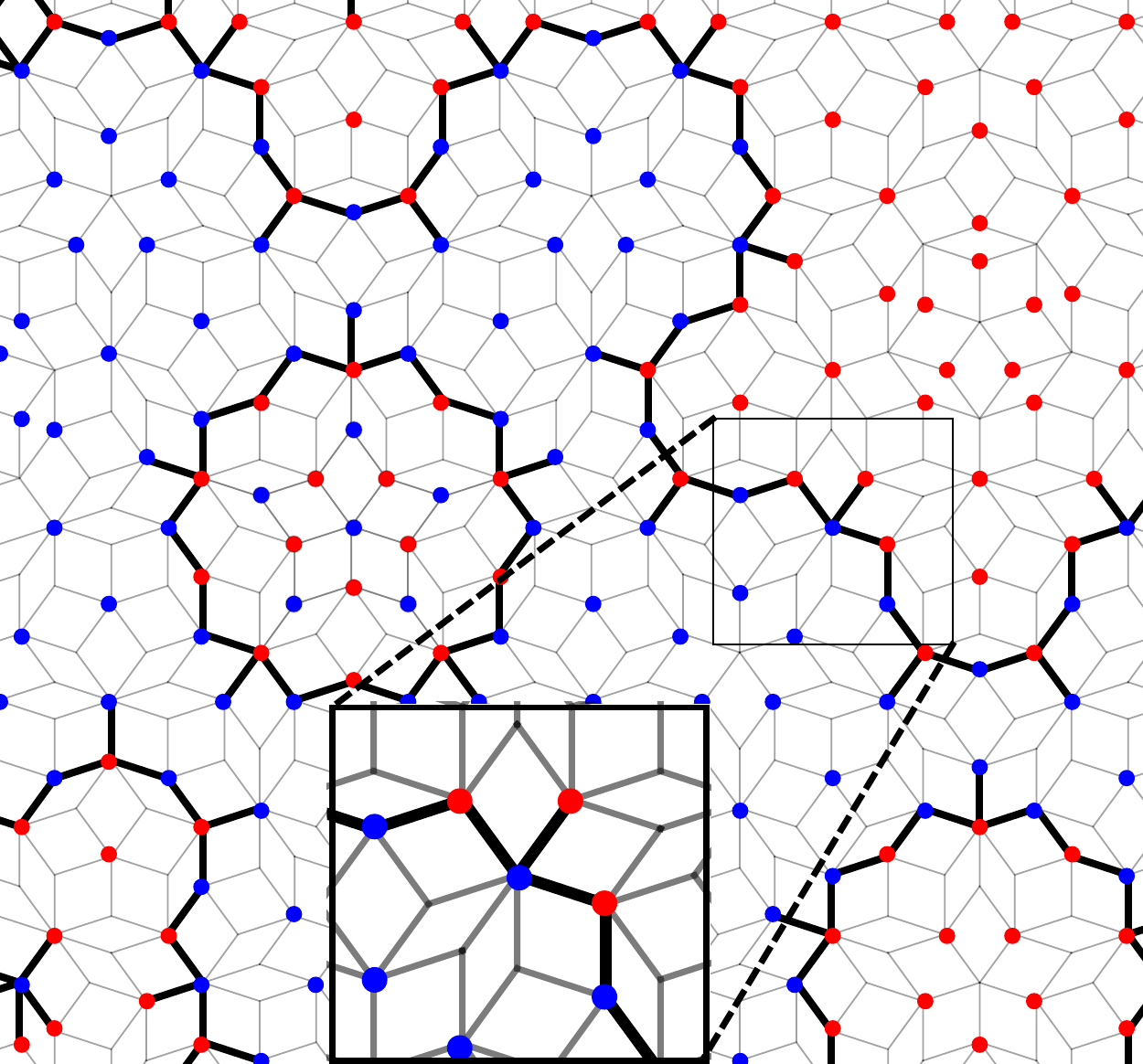}
\caption{Excluded sites for a portion of the rhombus lattice. They are colored red/blue based on which of the two global sublattices they belong to. Bold links,
connecting red and blue excluded sites, constitute boundaries of the domains.  }
\label{fig:rhombDecorated}
\end{figure}

To clarify the origin of the zero energy states, we again mark all sites for which the wave-function of the zero energy states vanishes (we call them excluded sites), shown
Fig.~\ref{fig:rhombDecorated}. Unlike the kite-and-dart lattice, the excluded sites do not occupy a single sublattice. They are 
marked in red if they belong to A sublattice and in blue if they belong to B. One notices that the two colors segregate into domains with well-defined boundaries, which run between alternating red and blue excluded sites (links marked as bold in Fig.~\ref{fig:rhombDecorated}). Specifically, if all the bold links are cut, 
the lattice segregates into isolated domains. Within each domain all the excluded sites are of the same color, i.e. they belong to the same (minority) sublattice. Note that, in the smallest domains there are a few accidental excluded sites on the majority sublattice. In the adjacent domain, all the excluded side again belong to one (minority) sublattice only, which is, however, the opposite sublattice from the previous domain.  

One may count the sublattice mismatch locally for each domain and add up their absolute values for the entire lattice, i.e. compute $\langle (-1)^{\mathcal{S}_{\bf r}} \Delta n(\bf r) \rangle$, where $\mathcal{S}_{\bf r}$ is $\pm1$ for each domain. The result of that calculation is shown in Fig.~\ref{fig:rhombZESplot} as the ``local'' mismatch. It is clear that for large lattice sizes the local mismatch, defined this way, indeed accounts for the 
majority of the zero energy states. The problem, however,  is that the lattice is not cut across the bold links in Fig.~\ref{fig:rhombDecorated} and therefore the domains are actually coupled. One might expect thus that the coupling lifts the macroscopic degeneracy of the zero energy states.

To understand why the degeneracy is intact,  consider a Hamiltonian for two neighboring domains, grouping sites by the domain, $(1,2)$, and sublattice within a domain $(A,B)$, letting the $A$ sublattice be the minority one for each domain. As we will see later, this means that sublattices $A_1$ and $A_2$ are not on the same global sublattice. The corresponding Hamiltonian is
\begin{align}
	H' = 
	\bordermatrix{
	~   & A_1   & B_1 & A_2   & B_2 \cr
	A_1 & 0     & G_1 & C     & 0   \cr
	B_1 & G_1^T & 0   & 0     & 0   \cr
	A_2 & C^T   & 0   & 0     & G_2 \cr
	B_2 & 0     & 0   & G_2^T & 0
	}.
\end{align}
Each domain, $1,2$, is represented by a usual  bipartite lattice Hamiltonian, Eq.~(\ref{eq:off-diagonal}). The two are coupled by the $C$ term connecting solely the minority sites in domain $1$ to the minority sites in the domain $2$. This is because the bold links in Fig.~\ref{fig:rhombDecorated} connect only the minority lattice sites in both domains.  From this it is clear that if $\psi_0 = (0, \vec{x_1})^T$ is a zero eigenstate of the first domain, considered in isolation, then there  is a corresponding localized zero eigenstate of the whole system, $\Psi_0 = (0, \vec{x_1}, 0, 0)^T$. A similar analysis holds for the second domain. An equivalent way to see this is that, since the domains are only joined along the sites with no amplitude (excluded sites) for all zero energy states, the coupling cannot  perturb these states. 
This pattern also requires that the global sublattices, occupied with the excluded sites,  switch between the adjacent domains. 

Therefore, from the viewpoint of the zero energy states, the quasicrystal is partitioned into strictly disconnected domains.  One can thus choose a basis where 
all zero energy states are confined to the majority lattice of one of the domains. As a result, counting the {\em local} sublattice mismatch (i.e. inside each domain), $\langle (-1)^{\mathcal{S}_{\bf r}} \Delta n(\bf r) \rangle$, gives an accurate count of the zero energy states, as shown in Fig.~\ref{fig:rhombZESplot}. It is important to stress that the ``domain walls'' are impenetrable only for the zero energy states. All other states ($90\%$) are not localized to the domains and propagate freely between them.  

Looking at the large scale structure of the rhombus lattice one can find domains with sizes  at all  scales. Figure \ref{fig:nested-domains} shows a larger portion of the lattice divided into its domains. These have been colored red/blue based on which (global) sublattice the excluded sites occupy in that domain. There are domains contained in other, larger domains; looking at increasingly large portions of the lattice, these domains continue to arbitrarily large length scales. As one can see, there is an element of self-similarity in this picture \cite{nine_percent,HATAKEYAMA198779,exact-number,PhysRevB.95.024509,dimers}. 

Therefore, a theory that explains the fraction of zero energy states must be capable of explaining this self-similarity structure of the domain shapes and sizes. 
In Section \ref{sec:Calculation}, we will develop a real space RG treatment of the Penrose lattice growth, based on its inflation property\cite{PhysRevB.34.3904}. The number of inflation generations serves  as the RG ``time''.  We find that this treatment is capable of accurately predicting the number of domains and their size distribution.

\section{Perturbations of the zero energy states} \label{sec:robustness}

\subsection{Robustness to perturbations}  

\begin{figure}[htp]
\centering
\includegraphics[width=0.4\textwidth]{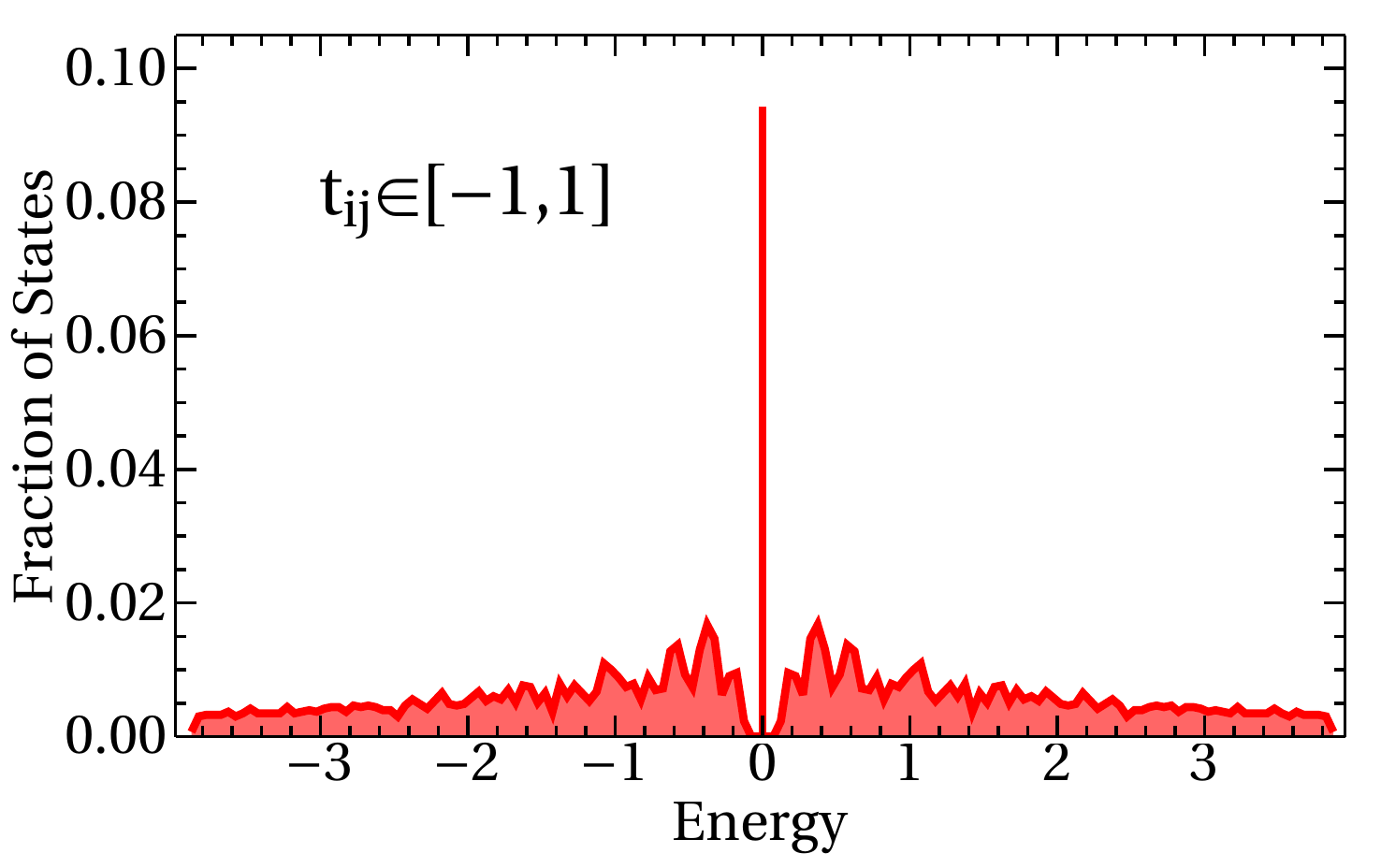}
\caption{DOS for a Penrose lattice with random nearest-neighbor hopping amplitudes drawn uniformly from $t_{ij} \in [-1,1]$ (one realization).}
\label{fig:random-t}
\end{figure}

\begin{figure}[htp]
\centering
\includegraphics[width=0.4\textwidth]{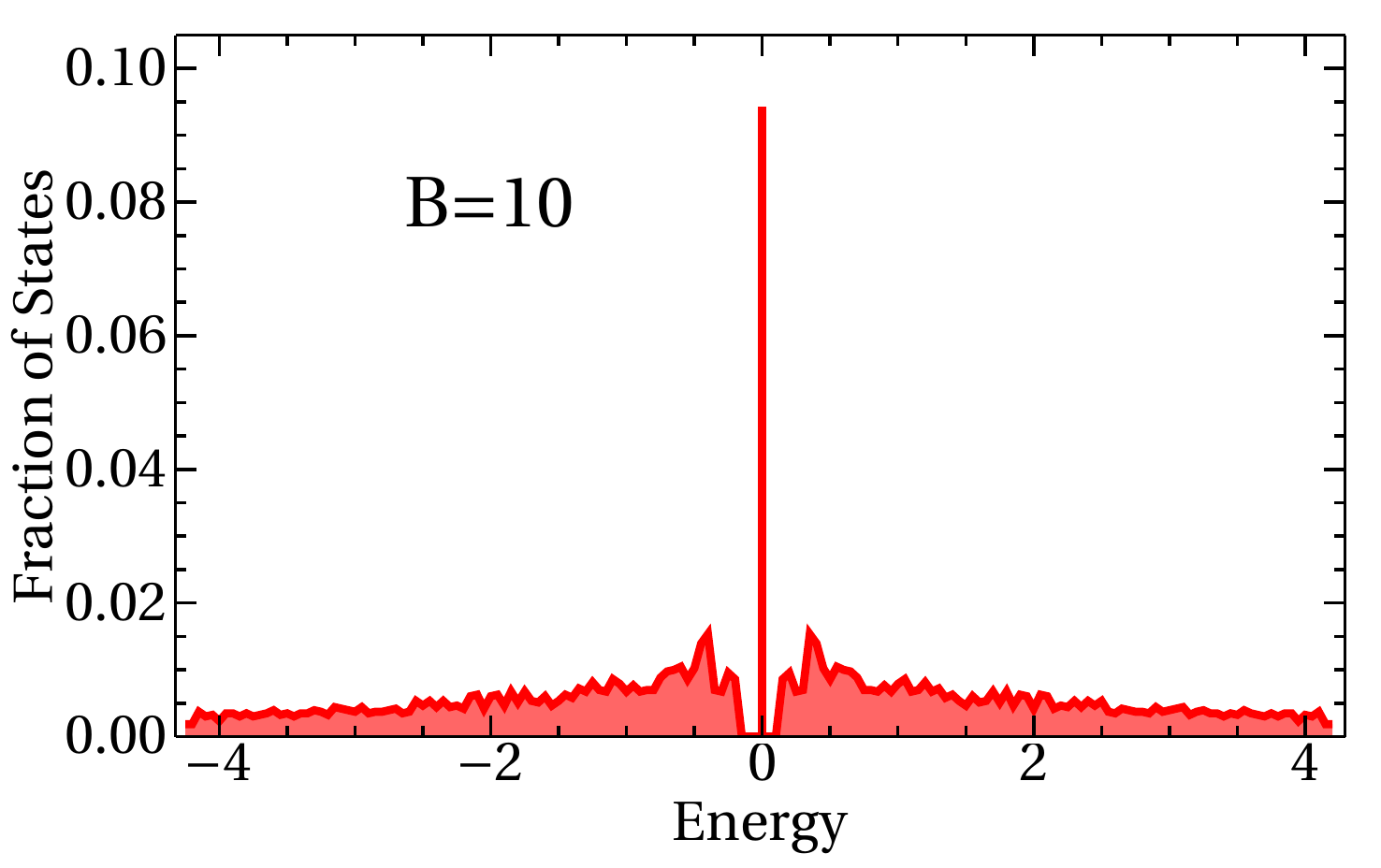}
\caption{DOS for a Penrose lattice in the presence of a perpendicular magnetic field of half flux quanta per small rhombus. }
\label{fig:magnetic-field}
\end{figure}

The existence of the zero energy states appears to be completely, or partially robust against a number of perturbations of the initial model. 
Figure \ref{fig:random-t} shows DOS of the Penrose lattice with random nearest neighbor hopping, $t$, drawn from a box distribution, $t\in [-1,1]$.
The number of zero energy states is the same as in the constant $t$ model.  This is not surprising, since both the domain structure and the mismatch count 
rely only on the geometry of the lattice, but not on specific hopping amplitudes. The same reasoning explains why a perpendicular magnetic field does not change the number of zero energy states, Fig.~\ref{fig:magnetic-field}. Indeed, the magnetic field enters as complex phases of the hopping amplitudes. Notice that both random hopping and the magnetic field do affect the DOS of non-zero energy states.   
  
Slightly less obvious is the effect of random dilution of the lattice by removing random sites. We found that this leads to a slow suppression of 
the number of the zero energy states, but not to their immediate disappearance.  This is also easy to understand, since the removal of the sites does not affect domain partitioning, but only removes sites from the interior or the boundary of a domain. Depending on whether a minority or a majority sublattice site is removed, the mismatch increases or decreases by one, adding or removing a zero energy state to/from the domain. Since it is $\approx 10\%$ more likely to remove a majority site, there is a slight tendency towards decreasing the number of zero energy states upon dilution.

\begin{figure}[htp]
\centering
\includegraphics[width=0.45\textwidth]{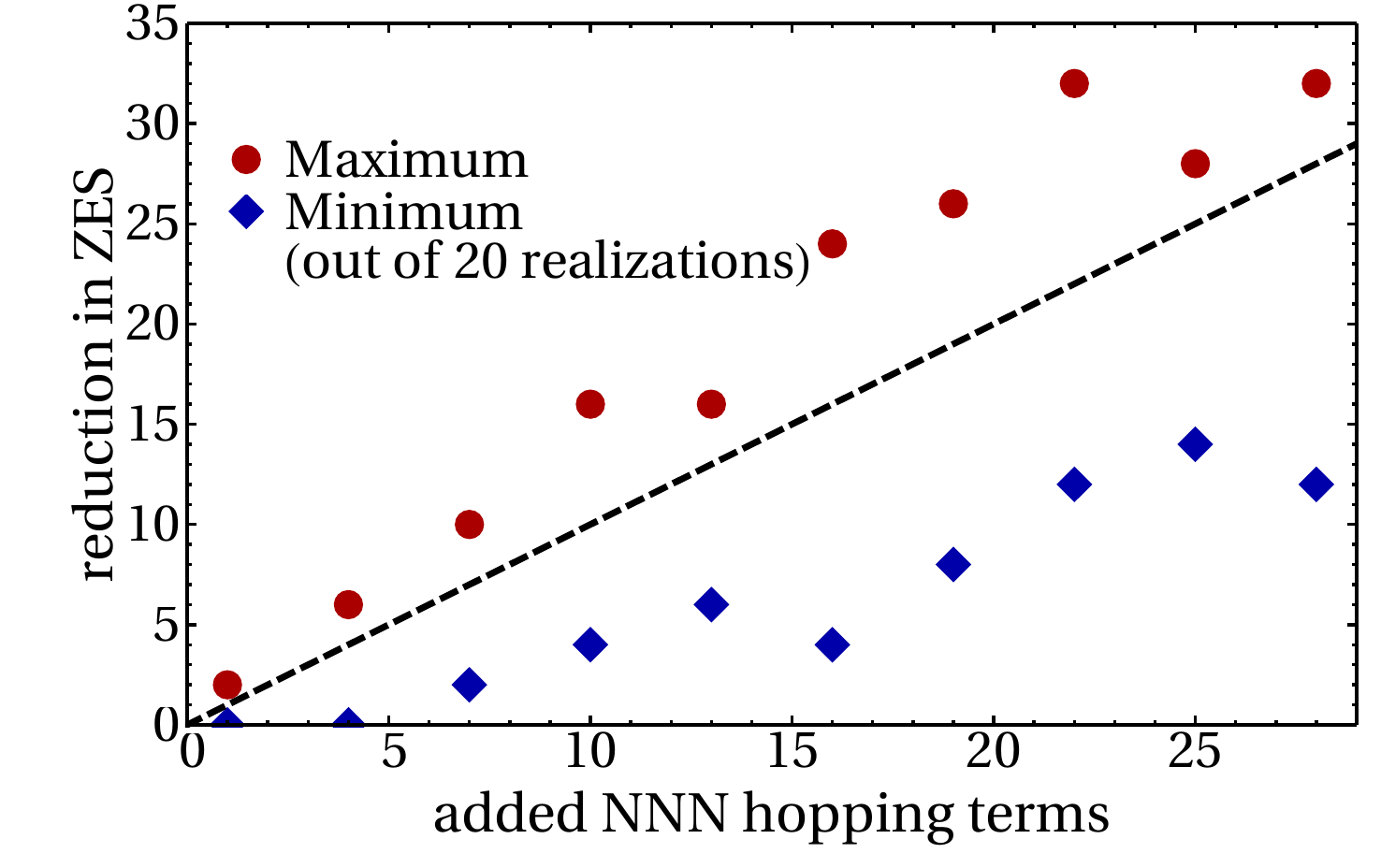}
\caption{Number of the zero energy states (ZES) removed upon addition of random NNN links. A site was randomly chosen, and a hopping amplitude was added to a random NNN of this site. Twenty realizations were considered; here, we plot only the maximal and minimal reductions of zero energy states. The dashed line has slope one, and is just a guide for the eyes. The lattice contains 4581 sites and 441 zero energy states.}  
\label{fig:reduction}
\end{figure}

The most severe perturbation is addition of next-nearest-neighbor (NNN) (i.e. along diagonals of some rhombuses) hopping amplitudes. Such a perturbation violates the bipartite nature of the lattice. If introduced across the domain boundary, it leads to inter domain coupling involving the majority sublattices. This could 
potentially eliminate all zero energy states in both domains. Yet this is not the case, as shown in Fig.~\ref{fig:reduction}, where we add NNN hopping in randomly chosen bonds. Each NNN link eliminates zero, one, or at most two zero energy states, depending on how many local majority sublattice sites it connects. Within a given domain one may choose a basis in the null  space, where all but one zero energy states have no amplitude at a given site of the majority sublattice. Thus, if this site is involved in NNN link, only 
the single state acquires a matrix element, which shifts it away from zero energy.  As a result, if less than   $\approx 20\%$ of sites participate in NNN links,
a fraction of the zero energy states persists.     

\subsection{Spatial structure of the zero energy states} 
\label{sec:spatial structure}

Since all zero energy states are exactly degenerate, one can choose any orthogonalized linear combination as a basis. One can always choose it to respect the 
domain structure, i.e. each state in the basis is fully localized within (the majority sublattice of) one domain and has zero overlap with states from all other domains. It is less straightforward to characterize individual states within a domain. To this end it is convenient to define the {\em basis-independent} projection matrix as
 \begin{equation}
 \label{eq:projection}
	P_{ij} = \sum_\alpha  \braket{i | \Psi^{(\alpha)}} \braket{\Psi^{(\alpha)} | j} , 
\end{equation}
where $\alpha$ labels zero energy states, $i,j$ label lattice sites, and $\Psi^{(\alpha)}$ is the wave-function corresponding to the $\alpha$ zero energy state. The projection matrix, $P_{ij}$, has block-diagonal form in the space of the domains with  non-zero elements only on the majority sublattice of the corresponding domain.  

Consider now a small local (i.e. at site $k$) perturbation of the onsite energy $V_{ij} = V \delta_{ik}\delta_{jk}$. In first-order degenerate perturbation theory, 
the energies of the zero energy states are shifted by the eigenvalues of the matrix  
 \begin{align}
 					\label{eq:V-perturbation}
	V^{\alpha\beta} = \braket{\Psi^{(\alpha)}|V|\Psi^{(\beta)}} = V \braket{\Psi^{(\alpha)} | k}   \braket{k | \Psi^{(\beta)}}.
\end{align} 
This is a rank-1 matrix with all eigenvalues, but one, equal to zero. Its only non-zero eigenvalue, $\delta E$, and the corresponding eigenfunction $\Phi_i^{(k)}$ are given by 
\begin{equation}
\delta E=VP_{kk};\qquad \Phi_i^{(k)} = P_{ik}/\sqrt{P_{kk}}.
\end{equation}
This means that the local perturbation (if on the majority sublattice) shifts the energy of a {\em single} zero energy state. Both the energy shift and the 
corresponding eigenfunction $\Phi_i^{(k)}$, localized around site $k$, are given in terms of $k$-th column of the projection matrix $P_{ik}$.  Notice that 
$\sum_k P_{kk}=$ {\em number of zero energy states}. As a result, the average energy shift susceptibility is $\overline{\delta E}/V = 2f/(1+f)\approx 0.179$, where 
$f=0.098$ is the fraction of zero energy states and the averaging is performed over the majority sublattice.  

\begin{figure}
\centering
\includegraphics[width=0.45\textwidth]{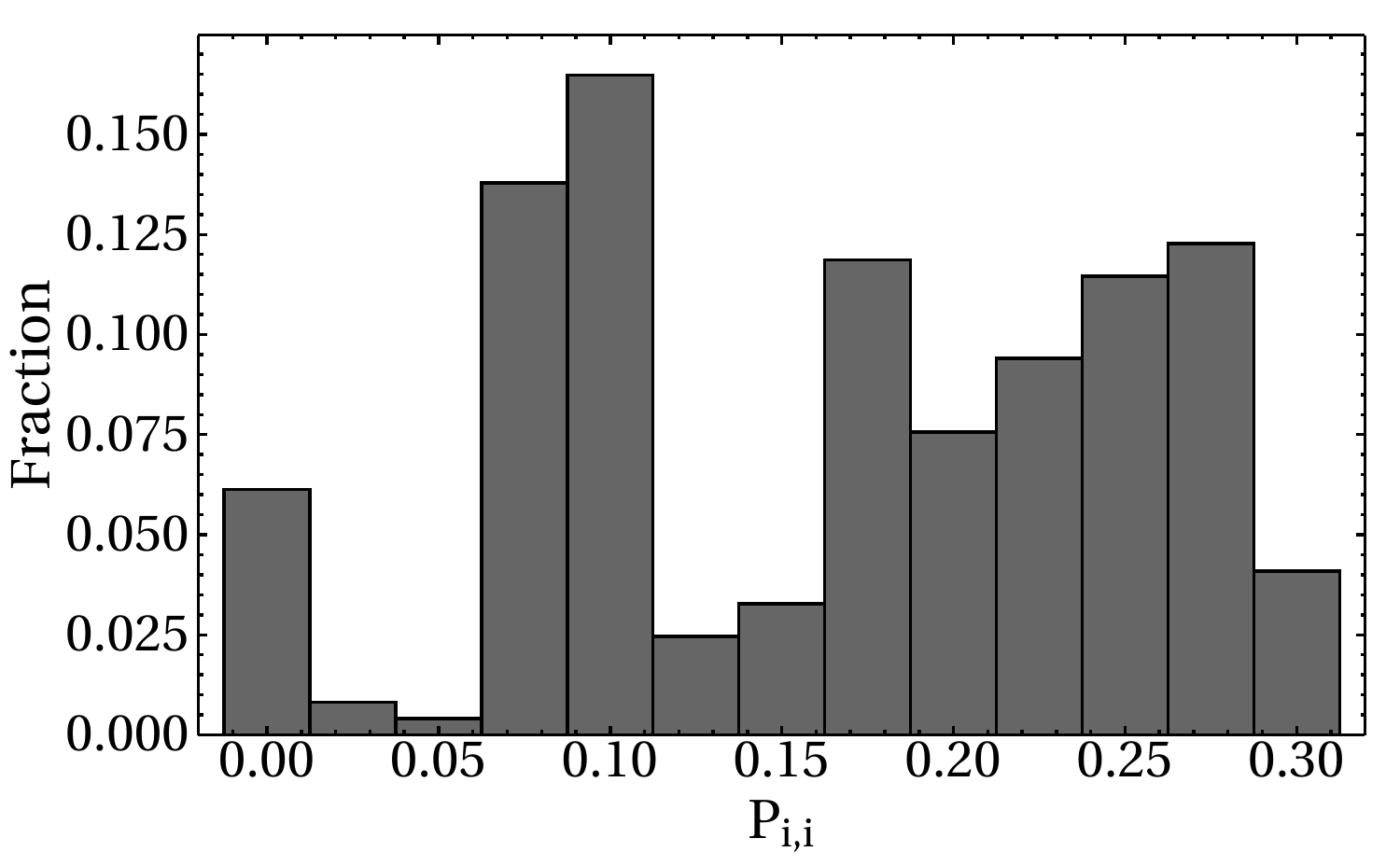}
\caption{Histogram of the diagonal entries of the projection matrix, $P_{ij}$, which determines the perturbative energy shift due to a local onsite potential (only majority sublattice is kept).}
\label{fig:PiiHistogram}
\end{figure}

\begin{figure}
\centering
\includegraphics[width=0.5\textwidth]{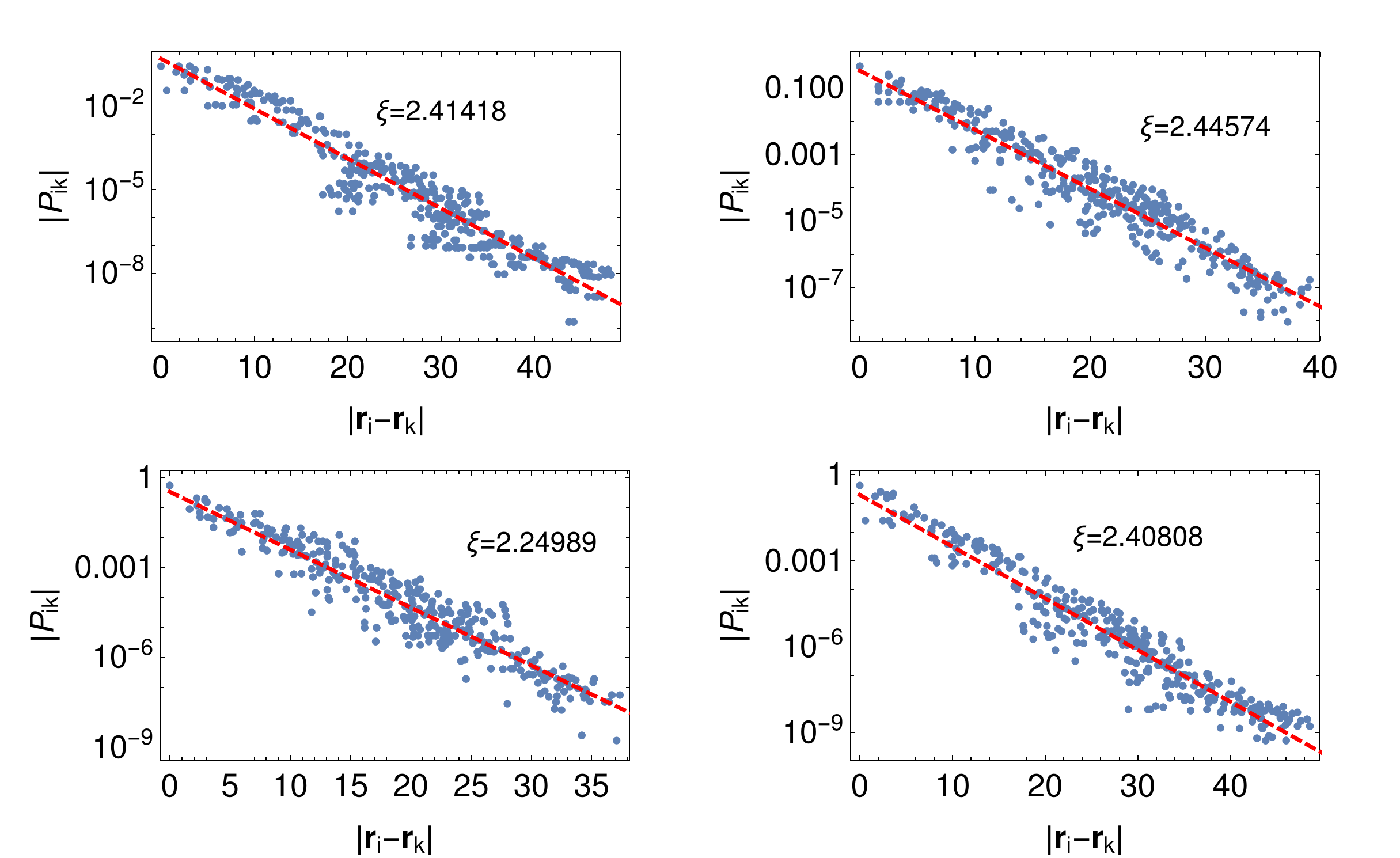}
\caption{Wave-functions of the perturbed states for perturbations on four different sites. The horizontal axis is the distance from the perturbation site in units of the rhombus edge length. The best fit localization lengths, $\xi_k$, are shown.}
\label{fig:decays}
\end{figure}

Figure \ref{fig:PiiHistogram} shows the distribution of the diagonal elements of the projection matrix, $P_{ii}$ , for the lattice with 4581 sites, colored in Fig.~\ref{fig:nested-domains}, which contains 441 zero energy states.  As expected, the mean susceptibility is about $0.18$, while the standard deviation is $0.085$. Figure 
\ref{fig:decays} shows log-plots of $P_{ik}\propto \Phi_i^{(k)}$ as a function of the distance $|\mathbf{r}_i - \mathbf{r}_k|$ in units of the rhombus side. Four randomly chosen sites $k$ are shown, while {\em all} other sites exhibit the same pattern.  It is evident that $\Phi_i^{(k)}\approx \sqrt{P_{kk}}\exp\left[ -|\mathbf{r}_i - \mathbf{r}_k|/\xi_k \right]$, where the localization length $\xi_k=2.4\pm 0.2$. We conclude thus that local perturbations of the zero energy manifold lead to a spatially localized response within the given domain.

\section{Real Space RG} \label{sec:RG}
\label{sec:Calculation}

\subsection{Background} 
\begin{figure}
\centering
\includegraphics[width=.45\textwidth]{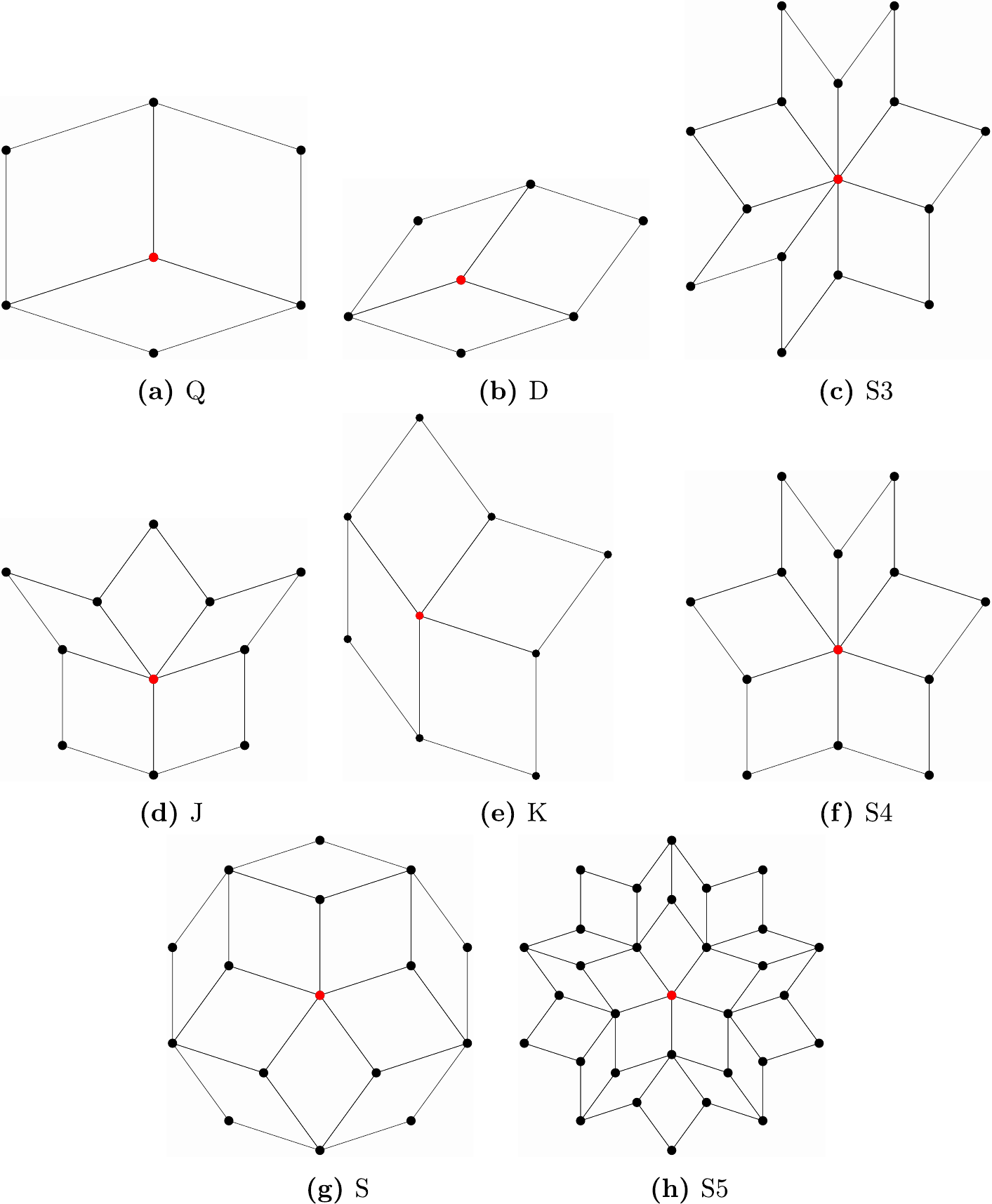}
\caption{The neighborhoods associated with each of the eight distinct types of sites. Each neighborhood uniquely identifies the central red site.}
\label{fig:neighborhoods}
\end{figure}

We now turn to develop a method to find the local sublattice mismatch and the number of domains in lattices of progressively increasing sizes. There are only eight distinct site types in the Penrose rhombus lattice. We follow the standard notations given in e.g. Ref.~\onlinecite{de1981ned}, which are specified in Fig. \ref{fig:neighborhoods}. Therefore a finite patch of the lattice can be characterized by an eight-dimensional vector, $\vec n$, whose entries correspond to the number of sites of a given type
\begin{align}
	\vec n = \begin{pmatrix}Q & D & S3 & J & K & S4 & S & S5\end{pmatrix}^T. 
\end{align}

The key to our approach is the {\em inflation property} of the Penrose lattice \cite{PhysRevB.34.3904}. This is a set of partition rules for all  rhombuses that generates another valid patch of the lattice with more sites. Since the rules are local and specific to a given type of the vertex, they can be represented as a matrix acting on the vector $\vec n$. In the bulk, this matrix is given by
\begin{align}                \label{eq:M-matrix}
	M = 
\begin{pmatrix}
 0 & 2 & 3 & 1 & 1 & 4 & 5 & 0 \\
 1 & 0 & 0 & 0 & 0 & 0 & 0 & 0 \\
 0 & 1 & 0 & 0 & 0 & 0 & 0 & 0 \\
 \frac{2}{3} & \frac{1}{3} & 0 & \frac{2}{3} & 1 & 0 & 0 & \frac{5}{3} \\
 0 & 0 & 0 & 1 & 0 & 0 & 0 & 0 \\
 0 & 0 & 0 & 0 & 1 & 0 & 0 & 0 \\
 0 & 0 & 0 & 0 & 0 & 0 & 0 & 1 \\
 0 & 0 & 1 & 0 & 0 & 1 & 1 & 0 \\
\end{pmatrix}.
\end{align}
The fractional numbers reflect the fact that a given site may be shared by several adjacent neighborhoods. This sharing is modified at the boundaries of the lattice. Thus the boundaries require some care and are dealt with in Appendix \ref{app:boundaries}. For now we proceed with the bulk of the lattice. 

Given the matrix $M$, one can compute the site counts after $k$ inflations with some initial state $\vec n_1$,
\begin{equation}    \label{eq:M-itirated}
	\vec n_k = M^k \vec n_1.
\end{equation}
The asymptotic growth rate of the number of sites as a function of the generation number is given by the matrix's largest eigenvalue $1+\tau$, where $\tau$ is the golden ratio. We can also find the relative distribution of site types in the infinite lattice as the corresponding eigenvector,
\begin{equation}
								\label{eq:site_prevalence} 
	\begin{pmatrix}
	{1\over \tau^{2}} & {1\over \tau^{4}} & {1\over \tau^{6}} & {1\over \tau^{3}} & {1\over \tau^{5}} & {1\over \tau^{7}} & {1\over \tau^6(1+\tau^2)} & 
	{1\over \tau^4(1+\tau^2)}
	\end{pmatrix},
\end{equation}
which reproduces the known distribution, see Ref. \onlinecite{site_fractions}. 

\begin{figure}
\centering
\includegraphics[width=0.3\textwidth]{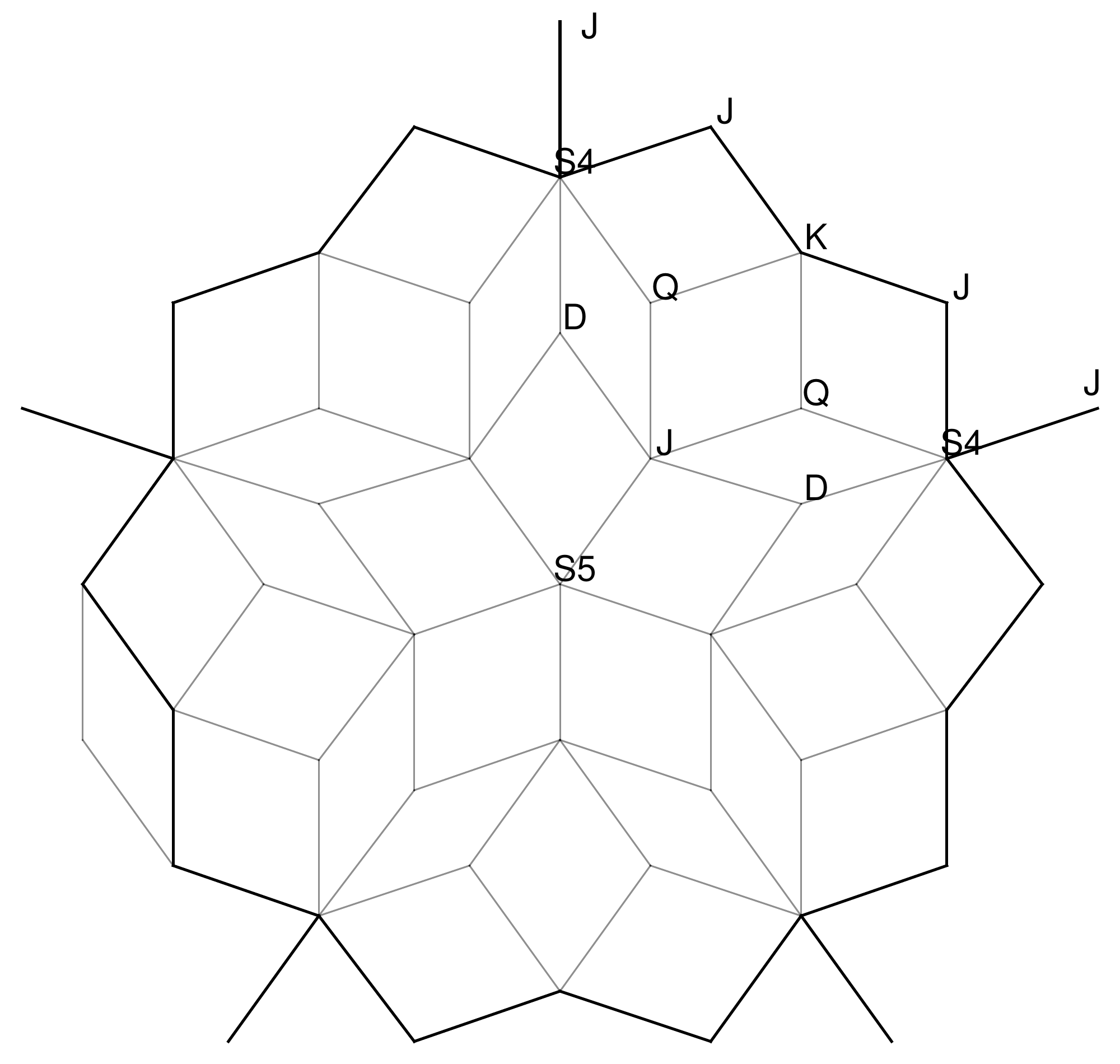}
\caption{A new emergent domain, obtained after two inflations of the $S5$ neighborhood. The sites are labeled according to the convention of Fig.~\ref{fig:neighborhoods}.  All other domains are results of the repeated inflations of this patch. It is also used as the initial seed to generate Penrose rhombus lattices with a larger number of sites via the inflation process.}		\label{fig:primdomain}
\end{figure}

Looking at Fig.~\ref{fig:nested-domains}, there are a number of domains of different shapes and sizes. In fact, all are generated from repeated inflations of a single ``seed'' domain. For example, the seed domain may be chosen as depicted in Fig.~\ref{fig:primdomain}. Upon repeated inflation steps it generates self-contained domains, all bounded by a line of excluded sites. Choosing this starting configuration also simplifies properly accounting for the boundary, the details of which are dealt with in Appendix \ref{app:boundaries}. 

Computing the local mismatch requires two inputs. The first is the number of new domains generated at the $k$-th inflation step, $A_k$. These newly born domains with the shape of the initial seed, Fig.~\ref{fig:primdomain},  may be seen in Figs.~\ref{fig:rhombDecorated} and \ref{fig:nested-domains}. Inspection shows that {\em all} new domains originate from the $S5$ neighborhoods -- see Fig. \ref{fig:neighborhoods} --  after two inflation iterations. We thus find:
\begin{equation}
				\label{eq:A}
A_k = (\vec n_{k-2})_{S5}.				
\end{equation}
The second input is the global mismatch, $G_k$, of A and B sublattices  after $k$ generations.   To find it, one needs to double the counting vector $\vec n$ to keep track to which sublattice a given site belongs: $\vec n = (\vec n_A, \vec n_B)^T$. Correspondingly the $M$ matrix also becomes $16\times 16$, and is presented in Appendix \ref{app:sublattices}. The global mismatch is given by   
\begin{equation}
	G_k = \sum_{i\in A}(\vec n_k)_i - \sum_{i\in B} (\vec n_k)_i, 
\end{equation}
which may be either positive or negative.

\begin{figure}
\centering
\includegraphics[width=0.45\textwidth]{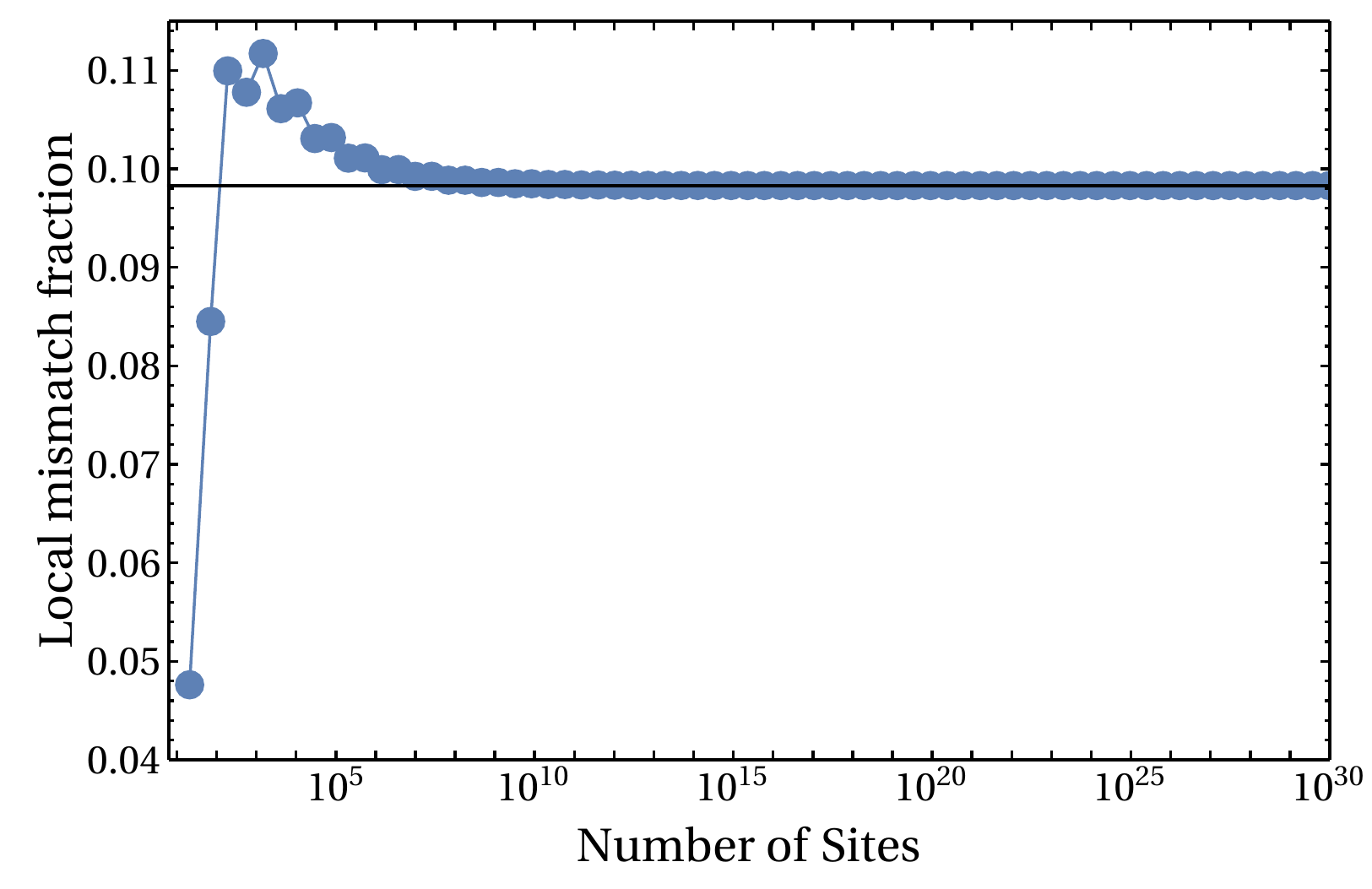}
\caption{Local mismatch fraction, $L_k/N_k$, as a function of the total number of sites after $k$ inflations, $N_k=\sum_i (\vec n_k)_i$.}
\label{fig:local-mismatch}
\end{figure}

With these two inputs, one can now evaluate the sum of all domain-specific local mismatches, $L_k$, after $k$ inflation steps. This is equivalent to the ``order parameter" $\langle (-1)^{\mathcal{S}_{\bf r}} \Delta n(\bf r) \rangle$ introduced above, and, as explained, is exactly the number of zero energy states in the DOS of the rhombus lattice. In Appendix \ref{app:L}, we show that it is given by: 
\begin{equation}\label{eq:final}
	L_k = G_k + 2 \sum_{l=1}^{k-1}A_{k-l}G_l. 
\end{equation} 
Equations (\ref{eq:M-matrix})--(\ref{eq:final}) provide  a complete iterative scheme to evaluate the number of zero energy states starting from any seed.  
It is straightforward to iterate  them to calculate the local mismatch for extremely large systems. Figure \ref{fig:local-mismatch} shows the result of such iteration for up to 70 generations with $\sim 10^{30}$ sites. After some initial fluctuations the mismatch as a fraction of total sites converges to $81-50\tau \approx 0.0983$.

We now proceed to derive this result analytically. The largest (non degenerate) eigenvalue of the $M$ matrix is $1+\tau$, signaling that the total number of sites, $N_k$, and the number of new domains, $A_k$, both scale as $(1+\tau)^k$. Thus, for $k\gg1$, with exponential accuracy: 
\begin{eqnarray}
N_k&=&\langle \vec 1|\psi_1\rangle (1+\tau)^k \langle\psi_1|\vec n_1\rangle,  \nonumber\\
A_k&=&\langle \vec S5|\psi_1\rangle (1+\tau)^{k-2} \langle\psi_1|\vec n_1\rangle, 
                       \label{eq:NkAk}
\end{eqnarray}
where $|\psi_1\rangle$ is the eigenvector corresponding to the largest eigenvalue, $\langle \vec 1|$ is the vector whose entries consist of $1$, and $\langle \vec S5|$ is a projection onto the $S5$ component. On the other hand, the global mismatch is given by $G_k=\langle \{\vec 1, -\vec 1\}|M^k|\vec n_1\rangle$, where $\bra{\{\vec 1, -\vec 1\}}$ has $1$ on entries corresponding to one sublattice and $-1$ on the other sublattice. It scales only as the second largest eigenvalue, since  $ \langle \{\vec 1, -\vec 1\}|\psi_1\rangle=0$.  The second largest eigenvalue is $2$ (note that the boundary has to be included to arrive at this number) and thus $G_k\propto 2^k$. 
Therefore, as already mentioned, $G_k\ll N_k, A_k,L_k$ for $k\gg 1$. Moreover, Eq.~(\ref{eq:final}) simplifies to
\begin{eqnarray}\label{eq:final1}
	L_k &\approx &   2 \sum_{l=1}^{k-1}A_{k-l}G_l =
	2\langle \vec S5|\psi_1\rangle  \langle\psi_1|\vec n_1\rangle \sum_{l=1}^{k-1}(1+\tau)^{k-l-2} G_l     \nonumber \\ 
	&=& 
	\frac{2\langle \vec S5|\psi_1\rangle  \langle\psi_1|\vec n_1\rangle }{(1+\tau)^2} \, (1+\tau)^k G(1/(1+\tau)), 
\end{eqnarray} 
where
\begin{eqnarray}\label{eq:Gsum}
	 G(1/(1+\tau)) = \sum_{l=1}^{\infty}(1+\tau)^{-l} G_l\equiv G,  
\end{eqnarray}
and we have extended the sum to infinity, since it is exponentially convergent.
The fraction of the zero energy states is thus     
 \begin{eqnarray}\label{eq:final2}
	\frac{L_k}{N_k} =
	\frac{\langle \vec S5|\psi_1\rangle  }{\langle \vec 1|\psi_1\rangle} \,  \frac{2 G}{(1+\tau)^2}. 
\end{eqnarray} 
The ratio  ${\langle \vec S5|\psi_1\rangle  }/\langle \vec 1|\psi_1\rangle= 1/ \tau^{4}(1+\tau^2)\approx 0.04$ is the global fraction of $S5$ sites, given by the last entry in Eq.~(\ref{eq:site_prevalence}). We thus obtain:
 \begin{eqnarray}\label{eq:final3}
	\frac{L_k}{N_k} =
	 \frac{2 G}{\tau^4 (1+\tau^2)(1+\tau)^2} =81-50\tau,   
\end{eqnarray} 
where we used that $2G=7+6\tau$, as evaluated in Appendix \ref{app:Gsum}. 

\begin{figure}
\centering
\includegraphics[width=0.45\textwidth]{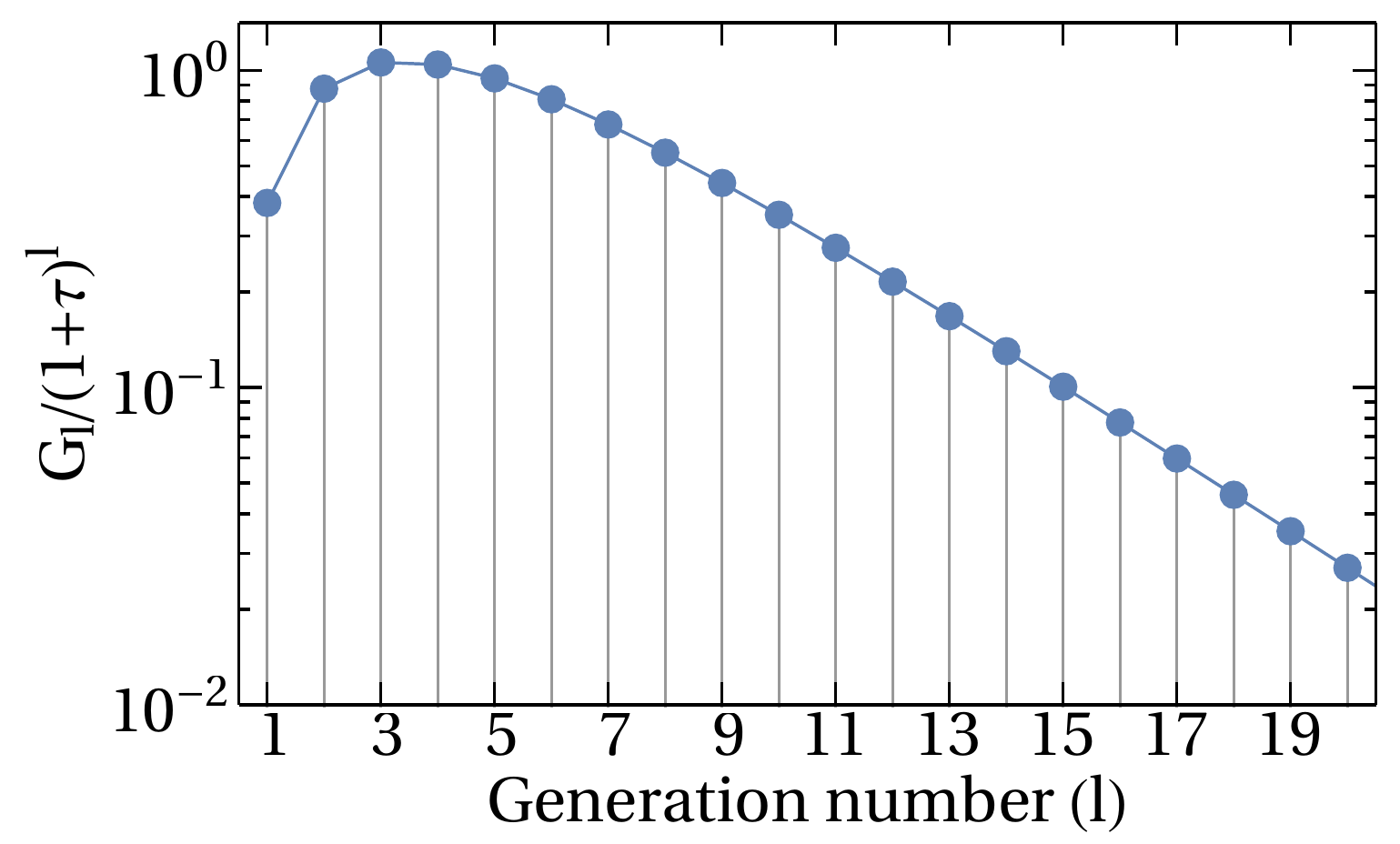}
\caption{Plot of $G_l/(1+\tau)^l$, the summand in Eq.~(\ref{eq:Gsum}), which is proportional to the relative contribution of a domain of size/age $l$ to the number of zero energy states.}
\label{fig:Gl}
\end{figure}

While this value agrees with previous reports \cite{Kohmoto_Sutherland,nine_percent,exact-number}, our approach provides interesting additional information. Figure~\ref{fig:Gl} show $\ln [(1+\tau)^{-l}G_l]$ as a function of the generation ``age" $l$.
According to Eq.~(\ref{eq:Gsum}), this quantity measures the relative contributions to the number of zero energy states of domains of ``age'' $l$, i.e. size $N_l$. The latter tells over how many lattice sites the corresponding zero energy states are extended. One notices that about half of the zero states are localized within ``young'' compact domains with $l\leq 5$. The remaining half falls within an exponential tail of larger ``older'' domains. For those we find a {\em scaling law} for the relative number of the zero energy states, $Z(N)$, extended over $N>N_5$ lattice sites: 
\begin{equation}
					\label{eq:scaling-law}
Z(N)\propto N^{-\eta};\qquad \eta= 1-\frac{\ln 2}{\ln(1+\tau)} \approx 	0.2798. 				
\end{equation}
Indeed, since $(1+\tau)$ and $2$ are the largest and the second largest eigenvalues of the $M$-matrix,  for $l\gtrsim 6$, one finds  $\ln Z=   \ln [(1+\tau)^{-l}G_l]\approx -l \ln[(1+\tau)/2]$. On the other hand, $\ln N = l \ln(1+\tau)$, from which Eq. (\ref{eq:scaling-law}) follows.

\section{Conclusions}
\label{sec:conclusions}

We have discussed the nature of the zero energy states in Penrose quasicrystals. We have shown that the lattice is subdivided into a nested structure 
of self-similar domains.  Upon inflation, the domains are inevitably born from every $S5$-neighborhood site and continue to grow indefinitely, while new domains 
appear inside older ones. The domain boundaries have a property of being impenetrable walls for the zero energy states (but not for any other states). 
As a result, each domain contributes a number to the total of zero energy states given by the mismatch between A and B sublattices within this domain. 
The mismatches alternate between successive domains, yielding no global mismatch. Yet, $\sim 10\%$ of all states are at exactly zero energy, due to the combined local mismatches in all the domains. The macroscopically degenerate zero energy states may be chosen to be localized within the respective domains.    

Utilizing the self-similar structure of the domains, we developed a real space RG evolution procedure, where the generation number, $k$, plays the same role as the 
RG ``time''. Notice that the lattice size $N_k$ grows exponentially with ``time'' as $N_k \sim(1+\tau)^k$. In other words, $k\sim \ln N_k$, as is common for 
real space RG. This procedure is capable of accurately counting domains, their sizes, and sublattices mismatch. It reproduces the $81-50\tau\approx 0.0983$ fraction of the zero energy states, derived before from different perspectives \cite{Kohmoto_Sutherland,nine_percent,exact-number}.   

The zero energy states and the domains supporting them appear to be robust against a number of perturbations. Random hopping and magnetic field do not alter their number at all. Random dilution of the lattice leads to a very slow decrease in their number. Finally, inclusion of NNN links, which violate the bipartite property of the original lattice, kills them with the average rate of a single state per NNN link. 

The robustness of the domain structure and the zero energy states inside the domains raises the question if they are of a topological origin. We have not been able
to find convincing arguments for or against this premise.   One tantalizing observation is that the Penrose lattice is a cross-section of a five dimensional (5D) 
cubic crystal. The latter has a bipartite structure, which is directly inherited by the Penrose tiling. The bipartite hopping Hamiltonian belongs to BD1 Altland-Zirnbauer symmetry class \cite{Ryu2016}, which is topological in 5D with the $Z$ homotopy group. One may thus wonder if the domains and zero energy states 
may be a legacy of their 5D topological parent.

\begin{acknowledgments} 
We thank F. Burnell and H. Manoharan for fruitful discussions. EDR and RMF were supported by the National Science
Foundation through the UMN MRSEC under DMR-1420013. AK was supported by  NSF grant DMR-1608238.
\end{acknowledgments}

\appendix
\section{Inflation procedure with sublattices}
\label{app:sublattices}

To calculate the global mismatch we need to track the sublattice that each site belongs to. We double the length of the $\vec n$ vector that counts sites, now tracking $A$ sites and $B$ sites separately.
\begin{align*}
\setcounter{MaxMatrixCols}{16}
	\vec n &= \begin{pmatrix}\vec{n}_A & \vec{n}_B \end{pmatrix}^T \\
	&= \begin{pmatrix}
	Q_{A} & D_{A} & S3_{A} & \hdots
	& Q_{B} & D_{B} & S3_{B} & \hdots
	 \end{pmatrix}
\end{align*}

Now we can write the inflation matrix in block form, in terms of whether the new sites are on the same ($AA$ subscript) or opposite ($AB$ subscript) sublattice as the original site.

\begin{align}
\label{eq:AB}
	M' = 
	\begin{pmatrix}
	M_{AA} & M_{AB} \\
	M_{AB} & M_{AA}
	\end{pmatrix}
\end{align}
That the diagonal blocks and off-diagonal blocks are the same comes from our freedom to choose which sublattice we label as $A$. To keep the total number of sites the same as in the original matrix we must have $M_{AA} + M_{AB} = M$. So it is sufficient to only give one block explicitly.

\begin{align}
\label{eq:AA}
M_{AA} = 
\begin{pmatrix}
 0 & 2 & 0 & 0 & 1 & 0 & 0 & 0 \\
 1 & 0 & 0 & 0 & 0 & 0 & 0 & 0 \\
 0 & 0 & 0 & 0 & 0 & 0 & 0 & 0 \\
 0 & \frac{1}{3} & 0 & 0 & 1 & 0 & 0 & \frac{5}{3} \\
 0 & 0 & 0 & 1 & 0 & 0 & 0 & 0 \\
 0 & 0 & 0 & 0 & 0 & 0 & 0 & 0 \\
 0 & 0 & 0 & 0 & 0 & 0 & 0 & 0 \\
 0 & 0 & 1 & 0 & 0 & 1 & 1 & 0 \\
\end{pmatrix}
\end{align}

\section{Boundary contributions}
\label{app:boundaries}

To account for boundary contributions, we will separately track the number of sites on the boundary and in the bulk. Then we can use different rules to evolve the boundary and add boundary effects on interior sites near the boundary.  For the starting geometry considered here, the boundary is only composed of four kinds of sites $J,K,S4,$ and $S3$. We can add the counts of these sites by appending them to the end of our vector $\vec{n}$, leaving it with 12 entries.
\begin{align}
	\vec n_1 = &\left(\begin{array}{cccccccccccc}
	10 & 5 & 0 & 5 & 0 & 0 & 1 & 0 & 15 & 5 & 5 & 0
	\end{array}\right)\\[-2ex]\nonumber
	&\hphantom{\left(\right.}\underbrace{\hphantom{\begin{array}{cccccccc}
	10 & 5 & 0 & 5 & 0 & 0 & 1 & 0\end{array}}}_{\text{previous vector}}
	\!\!\underbrace{\hphantom{\begin{array}{cccc}15 & 5 & 5 & 0\end{array}}}_{\text{boundary entries}}
\end{align}
The transition matrix will have the general structures
\begin{align}
	M_1 = \begin{pmatrix}
	M & B' \\
	0 & B\end{pmatrix}
\end{align}
Here $M$ is the same bulk matrix above, $B$ is the matrix describing the evolution of the boundary and $B'$ gives the contribution of the boundary sites to the bulk counts. We note that the details of $B$ and $B'$ depend on the geometry. Here the boundary is not exactly the geometrical boundary, but rather defined by the rings of excluded sites.

\begin{figure}[htp]
	\centering
	\includegraphics[width=.225\textwidth]{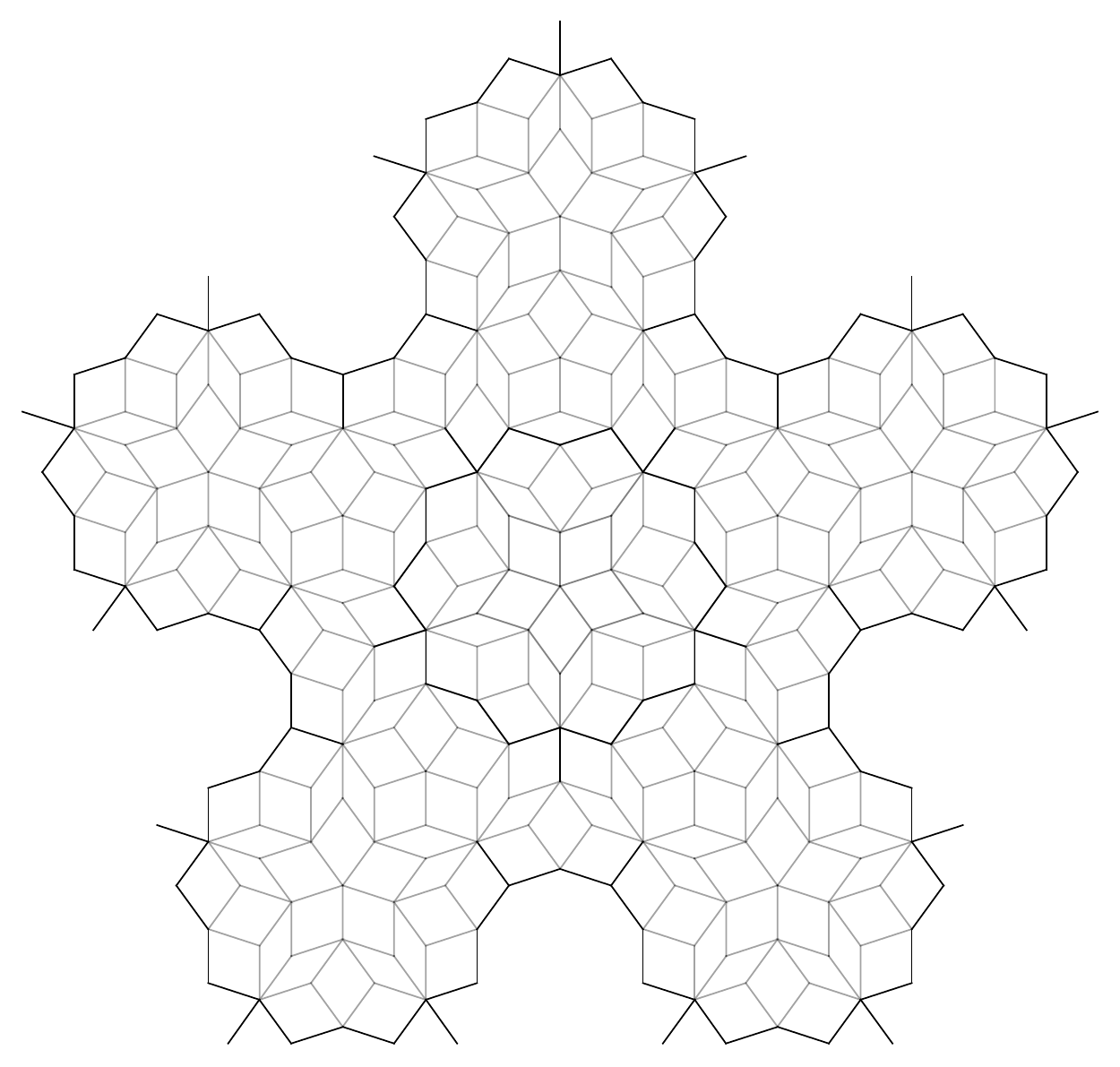}
	\includegraphics[width=.23\textwidth]{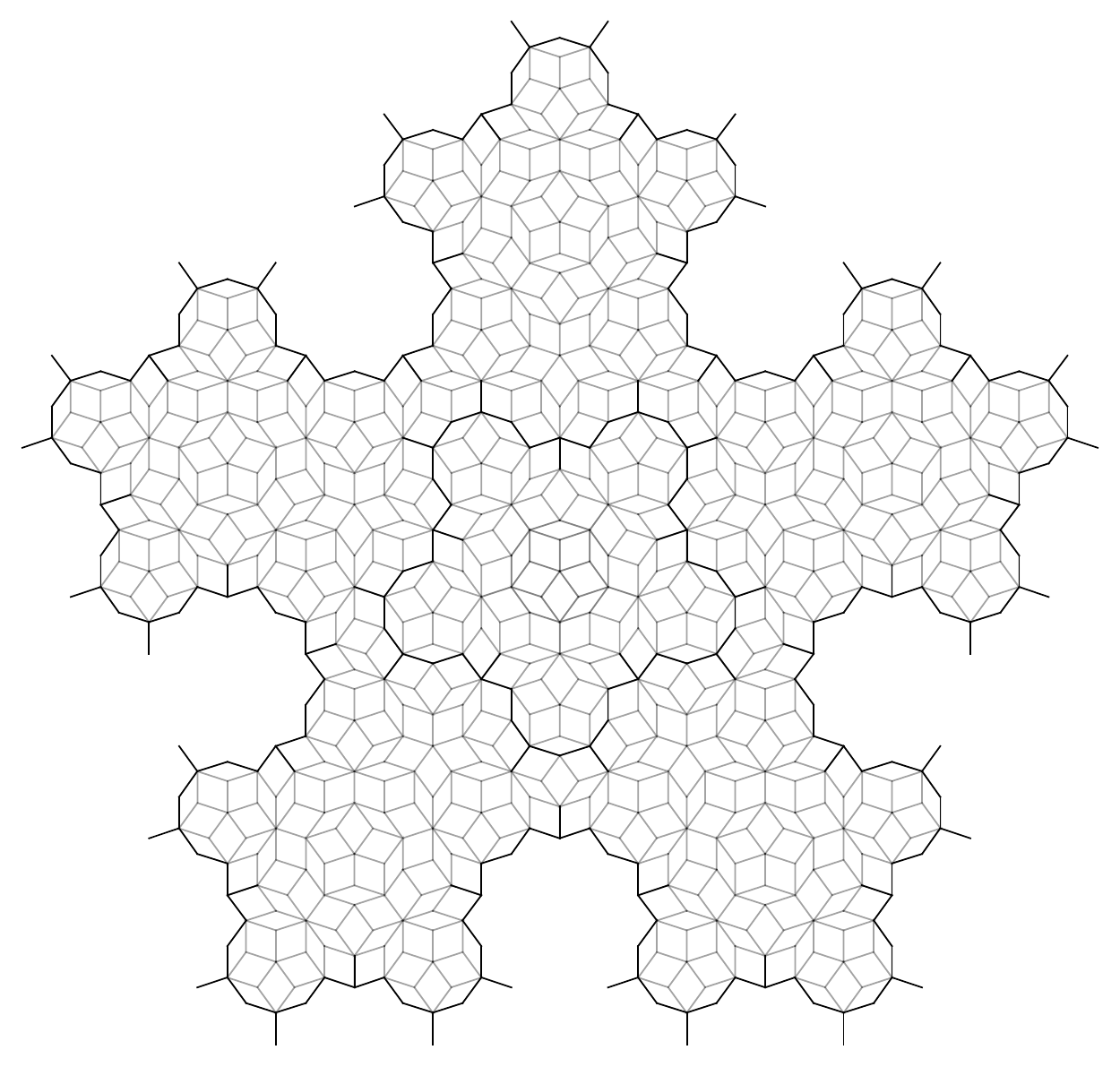}
	\caption{Boundaries for successive inflations. Note the S3 sites and J sites on inwards facing boundary on the left and right, respectively.}
	\label{fig:bd_parity}
\end{figure}

The main complication from adding the boundary is that the $B'$ matrix is not the same across inflations. Specifically, it alternates between two different versions. This is reflected in the structure of the boundary, since it alternates between two different sites bordering the interior, $S3$ and $J$ (see Figure \ref{fig:bd_parity}). The type of inward facing site changes the boundary effect on the interior. The matrices are explicitly given by

\begin{align}
	B = 
&\begin{pmatrix}
 1 & 1 & 1 & 1 \\
 1 & 0 & 0 & 0 \\
 0 & 1 & 0 & 0 \\
 1 & -1 & 0 & 0 \\
\end{pmatrix}
\\
B'_{\mathrm{even}} = 
&\begin{pmatrix}
 1 & 1 & 0 & 0 \\
 0 & 0 & 0 & 0 \\
 0 & 0 & -2 & 0 \\
 0 & 0 & -\frac{2}{3} & 0 \\
 0 & 0 & 0 & 0 \\
 0 & 0 & 0 & 0 \\
 0 & 0 & 0 & 0 \\
 0 & 0 & 0 & 0 \\
\end{pmatrix}
\quad
B'_{\mathrm{odd}} = 
\begin{pmatrix}
 0 & 0 & 4 & 3 \\
 0 & 0 & 0 & 0 \\
 0 & 0 & 0 & 0 \\
 0 & -\frac{4}{3} & 0 & 0 \\
 0 & 0 & 0 & 0 \\
 0 & 0 & 0 & 0 \\
 0 & 0 & 0 & 0 \\
 0 & 0 & 1 & 1 \\
\end{pmatrix}
\end{align}

The negative entries come from sites moving between the boundary and the bulk. In $B'_{\mathrm{even}}$ and $B'_{\mathrm{odd}}$, they come from sites being included on the inward facing ``spikes'' of the boundary.

These can also be extended to track the sublattice to which the site belong, as in Appendix \ref{app:sublattices}. All matrices have the structure of equation \ref{eq:AB} and obey the same constraint $M_{AA} + M_{AB} = M$ so we just give the $AA$ components, as in equation \ref{eq:AA}:

\begin{align}
	B_{AA} = 
	&\begin{pmatrix}
 0 & 1 & 1 & 1 \\
 1 & 0 & 0 & 0 \\
 0 & 0 & 0 & 0 \\
 1 & 0 & 0 & 0 \\
	\end{pmatrix}\\
	B'_{\mathrm{even},AA} =
	&\begin{pmatrix}
 0 & \frac{2}{2} & 0 & 0 \\
 0 & 0 & 0 & 0 \\
 0 & 0 & 0 & 0 \\
 0 & 0 & -\frac{2}{3} & 0 \\
 0 & 0 & 0 & 0 \\
 0 & 0 & 0 & 0 \\
 0 & 0 & 0 & 0 \\
 0 & 0 & 0 & 0 \\
	\end{pmatrix}
	\quad
	B'_{\mathrm{odd},AA} =
	\begin{pmatrix}
 0 & 0 & 0 & 0 \\
 0 & 0 & 0 & 0 \\
 0 & 0 & 0 & 0 \\
 0 & -\frac{4}{3} & 0 & 0 \\
 0 & 0 & 0 & 0 \\
 0 & 0 & 0 & 0 \\
 0 & 0 & 0 & 0 \\
 0 & 0 & 1 & 1 \\
	\end{pmatrix}
\end{align}

\section{Calculation of $G$}
\label{app:Gsum}
As just discussed, when accounting for the boundary, inflation behaves differently for even and odd generations. Thus, we must treat even and odd terms separately. Starting from our initial vector $\ket{\vec{n}_1}$, Eq. \ref{eq:M-itirated}, for an even number of inflations we can write
\begin{equation}
	\ket{\vec{n}_{2k+1}} = (M_{\mathrm{even}}M_{\mathrm{odd}})^k \ket{\vec{n}_1}
\end{equation}
We can expand in eigenvectors, $\{\ket i\}$, of ($M_{\mathrm{even}}M_{\mathrm{odd}}$). 
\begin{equation}
	\ket{\vec{n}_{2k+1}} = \sum_i \lambda_i^k w_i \ket i
\end{equation}
where $w_i$ are the weights connecting $\ket{\vec{n}_1}$ to $\ket i$. These are not $\braket{i|\vec{n}_1}$ because $(M_{\mathrm{even}}M_{\mathrm{odd}})$ is not symmetric and hence the $\ket i$ are not orthogonal. But nevertheless the weights exist and are unique.

The mismatch after an even number of inflations is
\begin{align}
	G_{2k+1} = \braket{\{\vec 1, \vec{-1}\}|\vec{n}_{2k+1}} = \sum_i \lambda_i^k w_i \braket{\{\vec 1, \vec{-1}\}|i}
\end{align}
From the product of the weights and the inner product, only 3 terms are non-zero, as listed in Table \ref{tab:Gconstants}.

\begin{table}
\begin{ruledtabular}
\begin{tabular}{l|l|l|l|l}
   $i$ & $w_i$ & $\lambda_i$ & $\braket{\{\vec 1, \vec{-1}\}|i}$ & $\braket{\{\vec 1, \vec{-1}\}|M_{\mathrm{odd}}|i}$ \\\hline
 2 & $\frac{10}{3}$    & 4   &  $\frac{18}{5}$ & $\frac{36}{5}$\\
 4 & $-\frac{\tau}{2}$   & $1+\tau$ &  $4+6\tau$ & $6+10\tau$\\
 13& $-\frac{1}{2} + \frac{\tau}{2}$ & $2-\tau$ & $10-6\tau$ & $16-10\tau$ \\
\end{tabular}
\end{ruledtabular}
\label{tab:Gconstants}
\caption{Parameters for the calculation of $G$}
\end{table}

For odd numbers of inflations we can define a similar expansion
\begin{equation}
	G_{2k+2} = \braket{\{\vec 1, \vec{-1}\}|n_{2k+2}} = \sum_i \lambda_i^k w_i \braket{\{\vec 1, \vec{-1}\}|M_{\mathrm{odd}}|i}
\end{equation}

Plugging in the summation for $G$, Eq. \ref{eq:Gsum}, we obtain:
\begin{align}
	\sum_{k=1}^\infty \frac{G_k}{(1+\tau)^k} 
	&= \sum_{k=0}^\infty \frac{G_{2k+1}}{(1+\tau)^{2k+1}} + \sum_{k=0}^\infty \frac{G_{2k+2}}{(1+\tau)^{2k+2}} \\
	&= \left(\frac{75}{44} + \frac{63}{44}\tau\right) + \left(\frac{91}{44} + \frac{57}{44}\tau\right)\\
	&= \frac{7}{2} + 3\tau \approx 8.3541
\end{align}

\section{Derivation of the equation for $L_k$}
\label{app:L}

To derive Eq.~(\ref{eq:final}), we start by noting that the smaller domains in Fig.~\ref{fig:nested-domains} look like the full lattice at an earlier stage of its evolution. 
We define the {\em exterior} region as the single domain that borders the boundary of the lattice. In Figure \ref{fig:nested-domains} this is the large blue region. Next we define the {\em top level} domains as domains that border this exterior region and hence are not contained inside anything except the exterior region. All red domains in Figure \ref{fig:nested-domains} are top level while only the small blue domain contained inside the central red domain is not top level. The number of {\em new} top level domains created within the exterior (outer blue region in Fig.~\ref{fig:nested-domains}) in the $k$-th generation is denoted by $T_k$.  We only count new top level domains because we can trivially find the number of older top level domains from this. A top level domain never stops being a top level, so the number of top level domains of age $l$ at a step $k$ is merely the number of domains that were newly created exactly $l$ generations ago, $T_{k-l}$.

Recall that the total number of new domains created at step $k$ is $A_k$. Since domains of a given age are identical, we know that a domain of age $l$ creates $A_l$ new domains inside of it. As just mentioned, the number of top level domains of age $l$ at step $k$ is $T_{k-l}$. Thus the number of new domains created inside domains of age $l$ is $A_l T_{k-l}$. We can sum this over $l$ to count all {\em non} top level domains created and subtract it from the total to find $T_k$, 
\begin{equation}\label{eq:T}
	T_k = A_k - \sum_{l=1}^{k-1}A_l T_{k-l}.
\end{equation}

Finally, we combine these relationships to find the total local mismatch, $L_k$, after $k$ generations. We can use a similar summation as Eq.~(\ref{eq:T}) to recursively count the local mismatch from enclosed domains. At generation $k$ there are $T_{k-l}$ top level domains of age $l$ and each contributes $L_l$ to the local mismatch. All is left is then to count the mismatch of the exterior region (the outer blue area in Figure \ref{fig:nested-domains}). We start with a simple example. 
\begin{figure}
\centering
\includegraphics[width=0.25\textwidth]{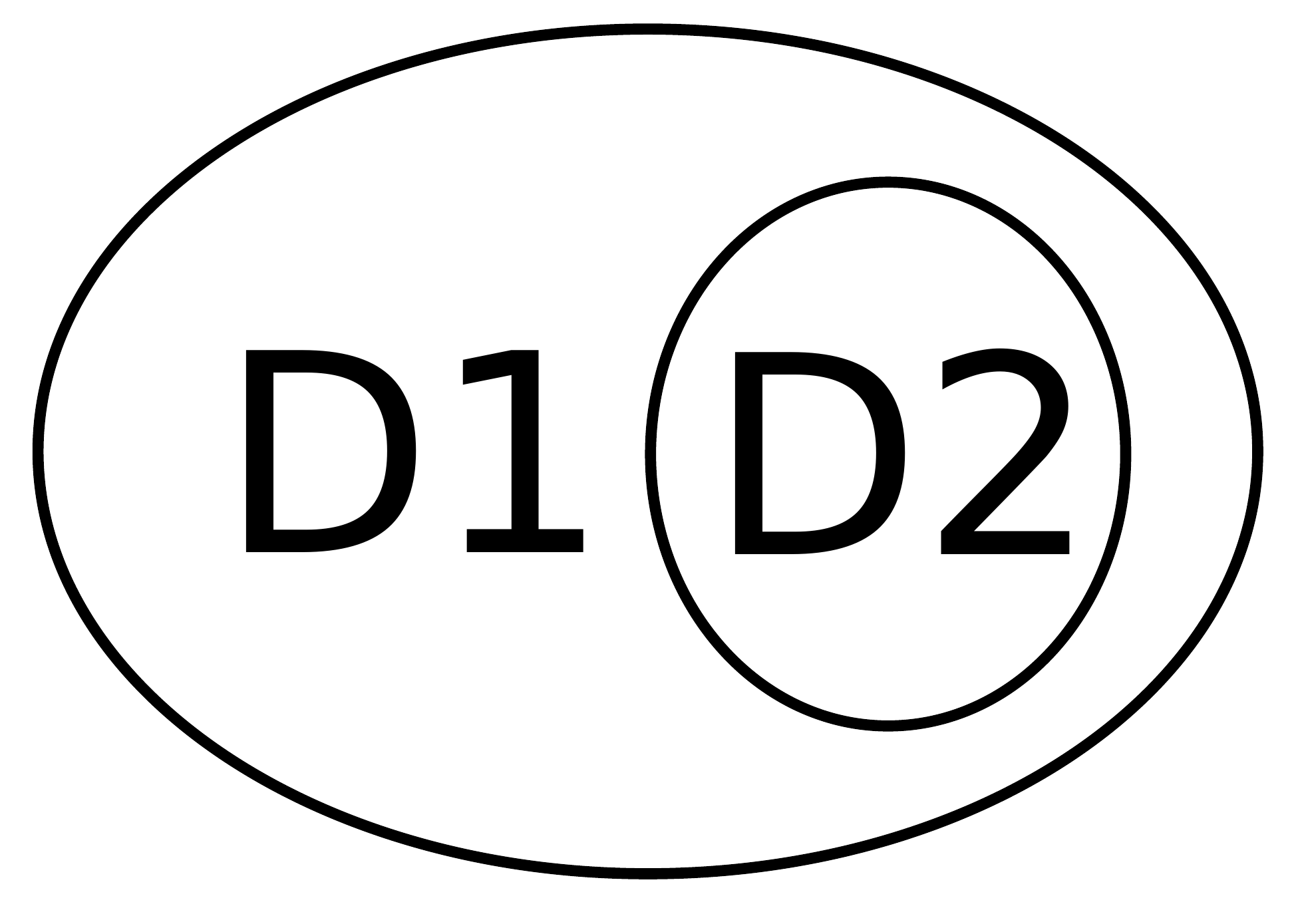}
\caption{Domain D1 contains another domain D2.}
\label{fig:simple-domains}
\end{figure}

Consider a domain $D1$ with a single domain $D2$ inside of it, as shown in Fig.~\ref{fig:simple-domains}. Let the combined domain have a mismatch $\Delta N$ (i.e greater number of A sites than B sites) and the interior domain $D2$ have a mismatch $-\Delta N_{D2}$ (i.e a greater number of B sites than A sites). It is negative because the majority sublattice switches when moving across the domain boundary. Then we have $\Delta N = \Delta N_{D1} - \Delta N_{D2}$, or $\Delta N_{D1} = \Delta N + \Delta N_{D2}$. Moreover, we have $T_{k-l}$ domains of age $l$, each of which has a mismatch of $G_l$. Summing these contributions gives the mismatch in the exterior domain, $G_k + \sum_l^{k-1} G_l T_{k-l}$. Combining this with the local mismatch from the enclosed domains, we obtain an equation for the total local mismatch of the lattice after $k$ inflations,
\begin{equation}\label{eq:L}
	L_k = G_k + \sum_{l=1}^{k-1} G_l T_{k-l} + \sum_{l=1}^{k-1} L_l T_{k-l}.
\end{equation}
Notice that both equations (\ref{eq:T}) and (\ref{eq:L}) are crucially based on the self-similarity property: domains of age $l$ are identical to the 
entire lattice at the $l$-th generation. 

The next goal is to eliminate $T_k$ from Eqs.~(\ref{eq:T}) and (\ref{eq:L}). To this end, we encode the discrete sequences into generating functions, e.g. 
\begin{align*}
	L(x) = \sum_{k=1}^\infty L_k x^k.
\end{align*}
Since $L_k$ grows exponentially as $\sim(1+\tau)^k$, the series has a finite radius of convergence, $|x|<(1+\tau)^{-1}$. This allows us to rewrite the summations as algebraic multiplications: 
\begin{align*}
	A(x)T(x) = \sum_k^\infty \left[ \sum_{l=1}^{k-1} A_l T_{k-l} \right] x^k.
\end{align*}
Equations (\ref{eq:T}) and (\ref{eq:L}) then become algebraic, $T=A -AT$ and $L=G+GT+LT$, where we omitted the argument $x$ for brevity. From the first, one finds 
$T=A/(1+A)$, and the second yields: $L=G(1+T)/(1-T)=G(1+2A)$. Matching the coefficients multiplying $x^k$ gives Eq.~(\ref{eq:final}). 

\nocite{*}
\bibliography{bib.bib}

\begin{thebibliography}{42}%
\makeatletter
\providecommand \@ifxundefined [1]{%
 \@ifx{#1\undefined}
}%
\providecommand \@ifnum [1]{%
 \ifnum #1\expandafter \@firstoftwo
 \else \expandafter \@secondoftwo
 \fi
}%
\providecommand \@ifx [1]{%
 \ifx #1\expandafter \@firstoftwo
 \else \expandafter \@secondoftwo
 \fi
}%
\providecommand \natexlab [1]{#1}%
\providecommand \enquote  [1]{``#1''}%
\providecommand \bibnamefont  [1]{#1}%
\providecommand \bibfnamefont [1]{#1}%
\providecommand \citenamefont [1]{#1}%
\providecommand \href@noop [0]{\@secondoftwo}%
\providecommand \href [0]{\begingroup \@sanitize@url \@href}%
\providecommand \@href[1]{\@@startlink{#1}\@@href}%
\providecommand \@@href[1]{\endgroup#1\@@endlink}%
\providecommand \@sanitize@url [0]{\catcode `\\12\catcode `\$12\catcode
  `\&12\catcode `\#12\catcode `\^12\catcode `\_12\catcode `\%12\relax}%
\providecommand \@@startlink[1]{}%
\providecommand \@@endlink[0]{}%
\providecommand \url  [0]{\begingroup\@sanitize@url \@url }%
\providecommand \@url [1]{\endgroup\@href {#1}{\urlprefix }}%
\providecommand \urlprefix  [0]{URL }%
\providecommand \Eprint [0]{\href }%
\providecommand \doibase [0]{http://dx.doi.org/}%
\providecommand \selectlanguage [0]{\@gobble}%
\providecommand \bibinfo  [0]{\@secondoftwo}%
\providecommand \bibfield  [0]{\@secondoftwo}%
\providecommand \translation [1]{[#1]}%
\providecommand \BibitemOpen [0]{}%
\providecommand \bibitemStop [0]{}%
\providecommand \bibitemNoStop [0]{.\EOS\space}%
\providecommand \EOS [0]{\spacefactor3000\relax}%
\providecommand \BibitemShut  [1]{\csname bibitem#1\endcsname}%
\let\auto@bib@innerbib\@empty
\bibitem [{\citenamefont {Shechtman}\ \emph {et~al.}(1984)\citenamefont
  {Shechtman}, \citenamefont {Blech}, \citenamefont {Gratias},\ and\
  \citenamefont {Cahn}}]{original}%
  \BibitemOpen
  \bibfield  {author} {\bibinfo {author} {\bibfnamefont {D.}~\bibnamefont
  {Shechtman}}, \bibinfo {author} {\bibfnamefont {I.}~\bibnamefont {Blech}},
  \bibinfo {author} {\bibfnamefont {D.}~\bibnamefont {Gratias}}, \ and\
  \bibinfo {author} {\bibfnamefont {J.~W.}\ \bibnamefont {Cahn}},\ }\href
  {\doibase 10.1103/PhysRevLett.53.1951} {\bibfield  {journal} {\bibinfo
  {journal} {Phys. Rev. Lett.}\ }\textbf {\bibinfo {volume} {53}},\ \bibinfo
  {pages} {1951} (\bibinfo {year} {1984})}\BibitemShut {NoStop}%
\bibitem [{\citenamefont {Ishimasa}\ \emph {et~al.}(1985)\citenamefont
  {Ishimasa}, \citenamefont {Nissen},\ and\ \citenamefont
  {Fukano}}]{PhysRevLett.55.511}%
  \BibitemOpen
  \bibfield  {author} {\bibinfo {author} {\bibfnamefont {T.}~\bibnamefont
  {Ishimasa}}, \bibinfo {author} {\bibfnamefont {H.-U.}\ \bibnamefont
  {Nissen}}, \ and\ \bibinfo {author} {\bibfnamefont {Y.}~\bibnamefont
  {Fukano}},\ }\href {\doibase 10.1103/PhysRevLett.55.511} {\bibfield
  {journal} {\bibinfo  {journal} {Phys. Rev. Lett.}\ }\textbf {\bibinfo
  {volume} {55}},\ \bibinfo {pages} {511} (\bibinfo {year} {1985})}\BibitemShut
  {NoStop}%
\bibitem [{\citenamefont {Wang}\ \emph {et~al.}(1987)\citenamefont {Wang},
  \citenamefont {Chen},\ and\ \citenamefont {Kuo}}]{PhysRevLett.59.1010}%
  \BibitemOpen
  \bibfield  {author} {\bibinfo {author} {\bibfnamefont {N.}~\bibnamefont
  {Wang}}, \bibinfo {author} {\bibfnamefont {H.}~\bibnamefont {Chen}}, \ and\
  \bibinfo {author} {\bibfnamefont {K.~H.}\ \bibnamefont {Kuo}},\ }\href
  {\doibase 10.1103/PhysRevLett.59.1010} {\bibfield  {journal} {\bibinfo
  {journal} {Phys. Rev. Lett.}\ }\textbf {\bibinfo {volume} {59}},\ \bibinfo
  {pages} {1010} (\bibinfo {year} {1987})}\BibitemShut {NoStop}%
\bibitem [{\citenamefont {Bendersky}(1985)}]{PhysRevLett.55.1461}%
  \BibitemOpen
  \bibfield  {author} {\bibinfo {author} {\bibfnamefont {L.}~\bibnamefont
  {Bendersky}},\ }\href {\doibase 10.1103/PhysRevLett.55.1461} {\bibfield
  {journal} {\bibinfo  {journal} {Phys. Rev. Lett.}\ }\textbf {\bibinfo
  {volume} {55}},\ \bibinfo {pages} {1461} (\bibinfo {year}
  {1985})}\BibitemShut {NoStop}%
\bibitem [{\citenamefont {Bindi}\ \emph {et~al.}(2009)\citenamefont {Bindi},
  \citenamefont {Steinhardt}, \citenamefont {Yao},\ and\ \citenamefont
  {Lu}}]{Bindi1306}%
  \BibitemOpen
  \bibfield  {author} {\bibinfo {author} {\bibfnamefont {L.}~\bibnamefont
  {Bindi}}, \bibinfo {author} {\bibfnamefont {P.~J.}\ \bibnamefont
  {Steinhardt}}, \bibinfo {author} {\bibfnamefont {N.}~\bibnamefont {Yao}}, \
  and\ \bibinfo {author} {\bibfnamefont {P.~J.}\ \bibnamefont {Lu}},\ }\href
  {\doibase 10.1126/science.1170827} {\bibfield  {journal} {\bibinfo  {journal}
  {Science}\ }\textbf {\bibinfo {volume} {324}},\ \bibinfo {pages} {1306}
  (\bibinfo {year} {2009})}\BibitemShut {NoStop}%
\bibitem [{\citenamefont {Kraus}\ \emph {et~al.}(2012)\citenamefont {Kraus},
  \citenamefont {Lahini}, \citenamefont {Ringel}, \citenamefont {Verbin},\ and\
  \citenamefont {Zilberberg}}]{topological-states}%
  \BibitemOpen
  \bibfield  {author} {\bibinfo {author} {\bibfnamefont {Y.~E.}\ \bibnamefont
  {Kraus}}, \bibinfo {author} {\bibfnamefont {Y.}~\bibnamefont {Lahini}},
  \bibinfo {author} {\bibfnamefont {Z.}~\bibnamefont {Ringel}}, \bibinfo
  {author} {\bibfnamefont {M.}~\bibnamefont {Verbin}}, \ and\ \bibinfo {author}
  {\bibfnamefont {O.}~\bibnamefont {Zilberberg}},\ }\href {\doibase
  10.1103/PhysRevLett.109.106402} {\bibfield  {journal} {\bibinfo  {journal}
  {Phys. Rev. Lett.}\ }\textbf {\bibinfo {volume} {109}},\ \bibinfo {pages}
  {106402} (\bibinfo {year} {2012})}\BibitemShut {NoStop}%
\bibitem [{\citenamefont {Apigo}\ \emph {et~al.}(2019)\citenamefont {Apigo},
  \citenamefont {Cheng}, \citenamefont {Dobiszewski}, \citenamefont {Prodan},\
  and\ \citenamefont {Prodan}}]{PhysRevLett.122.095501}%
  \BibitemOpen
  \bibfield  {author} {\bibinfo {author} {\bibfnamefont {D.~J.}\ \bibnamefont
  {Apigo}}, \bibinfo {author} {\bibfnamefont {W.}~\bibnamefont {Cheng}},
  \bibinfo {author} {\bibfnamefont {K.~F.}\ \bibnamefont {Dobiszewski}},
  \bibinfo {author} {\bibfnamefont {E.}~\bibnamefont {Prodan}}, \ and\ \bibinfo
  {author} {\bibfnamefont {C.}~\bibnamefont {Prodan}},\ }\href {\doibase
  10.1103/PhysRevLett.122.095501} {\bibfield  {journal} {\bibinfo  {journal}
  {Phys. Rev. Lett.}\ }\textbf {\bibinfo {volume} {122}},\ \bibinfo {pages}
  {095501} (\bibinfo {year} {2019})}\BibitemShut {NoStop}%
\bibitem [{\citenamefont {Zhao}\ \emph {et~al.}(2018)\citenamefont {Zhao},
  \citenamefont {Huo}, \citenamefont {Huang},\ and\ \citenamefont
  {Chen}}]{top-states-2}%
  \BibitemOpen
  \bibfield  {author} {\bibinfo {author} {\bibfnamefont {J.}~\bibnamefont
  {Zhao}}, \bibinfo {author} {\bibfnamefont {S.}~\bibnamefont {Huo}}, \bibinfo
  {author} {\bibfnamefont {H.}~\bibnamefont {Huang}}, \ and\ \bibinfo {author}
  {\bibfnamefont {J.}~\bibnamefont {Chen}},\ }\href {\doibase
  10.1002/pssr.201800322} {\bibfield  {journal} {\bibinfo  {journal} {physica
  status solidi (RRL) – Rapid Research Letters}\ }\textbf {\bibinfo {volume}
  {12}},\ \bibinfo {pages} {1800322} (\bibinfo {year} {2018})}\BibitemShut
  {NoStop}%
\bibitem [{\citenamefont {Dareau}\ \emph {et~al.}(2017)\citenamefont {Dareau},
  \citenamefont {Levy}, \citenamefont {Aguilera}, \citenamefont {Bouganne},
  \citenamefont {Akkermans}, \citenamefont {Gerbier},\ and\ \citenamefont
  {Beugnon}}]{PhysRevLett.119.215304}%
  \BibitemOpen
  \bibfield  {author} {\bibinfo {author} {\bibfnamefont {A.}~\bibnamefont
  {Dareau}}, \bibinfo {author} {\bibfnamefont {E.}~\bibnamefont {Levy}},
  \bibinfo {author} {\bibfnamefont {M.~B.}\ \bibnamefont {Aguilera}}, \bibinfo
  {author} {\bibfnamefont {R.}~\bibnamefont {Bouganne}}, \bibinfo {author}
  {\bibfnamefont {E.}~\bibnamefont {Akkermans}}, \bibinfo {author}
  {\bibfnamefont {F.}~\bibnamefont {Gerbier}}, \ and\ \bibinfo {author}
  {\bibfnamefont {J.}~\bibnamefont {Beugnon}},\ }\href {\doibase
  10.1103/PhysRevLett.119.215304} {\bibfield  {journal} {\bibinfo  {journal}
  {Phys. Rev. Lett.}\ }\textbf {\bibinfo {volume} {119}},\ \bibinfo {pages}
  {215304} (\bibinfo {year} {2017})}\BibitemShut {NoStop}%
\bibitem [{\citenamefont {Watanuki}\ \emph {et~al.}(2012)\citenamefont
  {Watanuki}, \citenamefont {Kashimoto}, \citenamefont {Kawana}, \citenamefont
  {Yamazaki}, \citenamefont {Machida}, \citenamefont {Tanaka},\ and\
  \citenamefont {Sato}}]{Watanuki12}%
  \BibitemOpen
  \bibfield  {author} {\bibinfo {author} {\bibfnamefont {T.}~\bibnamefont
  {Watanuki}}, \bibinfo {author} {\bibfnamefont {S.}~\bibnamefont {Kashimoto}},
  \bibinfo {author} {\bibfnamefont {D.}~\bibnamefont {Kawana}}, \bibinfo
  {author} {\bibfnamefont {T.}~\bibnamefont {Yamazaki}}, \bibinfo {author}
  {\bibfnamefont {A.}~\bibnamefont {Machida}}, \bibinfo {author} {\bibfnamefont
  {Y.}~\bibnamefont {Tanaka}}, \ and\ \bibinfo {author} {\bibfnamefont {T.~J.}\
  \bibnamefont {Sato}},\ }\href {\doibase 10.1103/PhysRevB.86.094201}
  {\bibfield  {journal} {\bibinfo  {journal} {Phys. Rev. B}\ }\textbf {\bibinfo
  {volume} {86}},\ \bibinfo {pages} {094201} (\bibinfo {year}
  {2012})}\BibitemShut {NoStop}%
\bibitem [{\citenamefont {Matsunami}\ \emph {et~al.}(2017)\citenamefont
  {Matsunami}, \citenamefont {Oura}, \citenamefont {Tamasaku}, \citenamefont
  {Ishikawa}, \citenamefont {Ideta}, \citenamefont {Tanaka}, \citenamefont
  {Takeuchi}, \citenamefont {Yamada}, \citenamefont {Tsai}, \citenamefont
  {Imura}, \citenamefont {Deguchi}, \citenamefont {Sato},\ and\ \citenamefont
  {Ishimasa}}]{PhysRevB.96.241102}%
  \BibitemOpen
  \bibfield  {author} {\bibinfo {author} {\bibfnamefont {M.}~\bibnamefont
  {Matsunami}}, \bibinfo {author} {\bibfnamefont {M.}~\bibnamefont {Oura}},
  \bibinfo {author} {\bibfnamefont {K.}~\bibnamefont {Tamasaku}}, \bibinfo
  {author} {\bibfnamefont {T.}~\bibnamefont {Ishikawa}}, \bibinfo {author}
  {\bibfnamefont {S.}~\bibnamefont {Ideta}}, \bibinfo {author} {\bibfnamefont
  {K.}~\bibnamefont {Tanaka}}, \bibinfo {author} {\bibfnamefont
  {T.}~\bibnamefont {Takeuchi}}, \bibinfo {author} {\bibfnamefont
  {T.}~\bibnamefont {Yamada}}, \bibinfo {author} {\bibfnamefont {A.~P.}\
  \bibnamefont {Tsai}}, \bibinfo {author} {\bibfnamefont {K.}~\bibnamefont
  {Imura}}, \bibinfo {author} {\bibfnamefont {K.}~\bibnamefont {Deguchi}},
  \bibinfo {author} {\bibfnamefont {N.~K.}\ \bibnamefont {Sato}}, \ and\
  \bibinfo {author} {\bibfnamefont {T.}~\bibnamefont {Ishimasa}},\ }\href
  {\doibase 10.1103/PhysRevB.96.241102} {\bibfield  {journal} {\bibinfo
  {journal} {Phys. Rev. B}\ }\textbf {\bibinfo {volume} {96}},\ \bibinfo
  {pages} {241102} (\bibinfo {year} {2017})}\BibitemShut {NoStop}%
\bibitem [{\citenamefont {Andrade}\ \emph {et~al.}(2015)\citenamefont
  {Andrade}, \citenamefont {Jagannathan}, \citenamefont {Miranda},
  \citenamefont {Vojta},\ and\ \citenamefont
  {Dobrosavljevi\ifmmode~\acute{c}\else \'{c}\fi{}}}]{heavy-fermion}%
  \BibitemOpen
  \bibfield  {author} {\bibinfo {author} {\bibfnamefont {E.~C.}\ \bibnamefont
  {Andrade}}, \bibinfo {author} {\bibfnamefont {A.}~\bibnamefont
  {Jagannathan}}, \bibinfo {author} {\bibfnamefont {E.}~\bibnamefont
  {Miranda}}, \bibinfo {author} {\bibfnamefont {M.}~\bibnamefont {Vojta}}, \
  and\ \bibinfo {author} {\bibfnamefont {V.}~\bibnamefont
  {Dobrosavljevi\ifmmode~\acute{c}\else \'{c}\fi{}}},\ }\href {\doibase
  10.1103/PhysRevLett.115.036403} {\bibfield  {journal} {\bibinfo  {journal}
  {Phys. Rev. Lett.}\ }\textbf {\bibinfo {volume} {115}},\ \bibinfo {pages}
  {036403} (\bibinfo {year} {2015})}\BibitemShut {NoStop}%
\bibitem [{\citenamefont {Matsukawa}\ \emph {et~al.}(2016)\citenamefont
  {Matsukawa}, \citenamefont {Deguchi}, \citenamefont {Imura}, \citenamefont
  {Ishimasa},\ and\ \citenamefont {Sato}}]{JPSJ.85.063706}%
  \BibitemOpen
  \bibfield  {author} {\bibinfo {author} {\bibfnamefont {S.}~\bibnamefont
  {Matsukawa}}, \bibinfo {author} {\bibfnamefont {K.}~\bibnamefont {Deguchi}},
  \bibinfo {author} {\bibfnamefont {K.}~\bibnamefont {Imura}}, \bibinfo
  {author} {\bibfnamefont {T.}~\bibnamefont {Ishimasa}}, \ and\ \bibinfo
  {author} {\bibfnamefont {N.~K.}\ \bibnamefont {Sato}},\ }\href {\doibase
  10.7566/JPSJ.85.063706} {\bibfield  {journal} {\bibinfo  {journal} {Journal
  of the Physical Society of Japan}\ }\textbf {\bibinfo {volume} {85}},\
  \bibinfo {pages} {063706} (\bibinfo {year} {2016})}\BibitemShut {NoStop}%
\bibitem [{\citenamefont {Sakai}\ \emph {et~al.}(2017)\citenamefont {Sakai},
  \citenamefont {Takemori}, \citenamefont {Koga},\ and\ \citenamefont
  {Arita}}]{PhysRevB.95.024509}%
  \BibitemOpen
  \bibfield  {author} {\bibinfo {author} {\bibfnamefont {S.}~\bibnamefont
  {Sakai}}, \bibinfo {author} {\bibfnamefont {N.}~\bibnamefont {Takemori}},
  \bibinfo {author} {\bibfnamefont {A.}~\bibnamefont {Koga}}, \ and\ \bibinfo
  {author} {\bibfnamefont {R.}~\bibnamefont {Arita}},\ }\href {\doibase
  10.1103/PhysRevB.95.024509} {\bibfield  {journal} {\bibinfo  {journal} {Phys.
  Rev. B}\ }\textbf {\bibinfo {volume} {95}},\ \bibinfo {pages} {024509}
  (\bibinfo {year} {2017})}\BibitemShut {NoStop}%
\bibitem [{\citenamefont {Ara\'ujo}\ and\ \citenamefont
  {Andrade}(2019)}]{Araujo19}%
  \BibitemOpen
  \bibfield  {author} {\bibinfo {author} {\bibfnamefont {R.~N.}\ \bibnamefont
  {Ara\'ujo}}\ and\ \bibinfo {author} {\bibfnamefont {E.~C.}\ \bibnamefont
  {Andrade}},\ }\href {\doibase 10.1103/PhysRevB.100.014510} {\bibfield
  {journal} {\bibinfo  {journal} {Phys. Rev. B}\ }\textbf {\bibinfo {volume}
  {100}},\ \bibinfo {pages} {014510} (\bibinfo {year} {2019})}\BibitemShut
  {NoStop}%
\bibitem [{\citenamefont {Zhang}\ \emph {et~al.}(2020)\citenamefont {Zhang},
  \citenamefont {Liu}, \citenamefont {Chen},\ and\ \citenamefont
  {Yang}}]{Zhang2020}%
  \BibitemOpen
  \bibfield  {author} {\bibinfo {author} {\bibfnamefont {Y.}~\bibnamefont
  {Zhang}}, \bibinfo {author} {\bibfnamefont {Y.-B.}\ \bibnamefont {Liu}},
  \bibinfo {author} {\bibfnamefont {W.-Q.}\ \bibnamefont {Chen}}, \ and\
  \bibinfo {author} {\bibfnamefont {F.}~\bibnamefont {Yang}},\ }\href@noop {}
  {\bibfield  {journal} {\bibinfo  {journal} {arXiv:2002.06485}\ } (\bibinfo
  {year} {2020})}\BibitemShut {NoStop}%
\bibitem [{\citenamefont {Cao}\ \emph {et~al.}(2020)\citenamefont {Cao},
  \citenamefont {Zhang}, \citenamefont {Liu}, \citenamefont {Liu},
  \citenamefont {Chen},\ and\ \citenamefont {Yang}}]{Cao2020}%
  \BibitemOpen
  \bibfield  {author} {\bibinfo {author} {\bibfnamefont {Y.}~\bibnamefont
  {Cao}}, \bibinfo {author} {\bibfnamefont {Y.}~\bibnamefont {Zhang}}, \bibinfo
  {author} {\bibfnamefont {Y.-B.}\ \bibnamefont {Liu}}, \bibinfo {author}
  {\bibfnamefont {C.-C.}\ \bibnamefont {Liu}}, \bibinfo {author} {\bibfnamefont
  {W.-Q.}\ \bibnamefont {Chen}}, \ and\ \bibinfo {author} {\bibfnamefont
  {F.}~\bibnamefont {Yang}},\ }\href@noop {} {\bibfield  {journal} {\bibinfo
  {journal} {arXiv:2001.07043}\ } (\bibinfo {year} {2020})}\BibitemShut
  {NoStop}%
\bibitem [{\citenamefont {Deguchi}\ \emph {et~al.}(2012)\citenamefont
  {Deguchi}, \citenamefont {Matsukawa}, \citenamefont {Sato}, \citenamefont
  {Hattori}, \citenamefont {Ishida}, \citenamefont {Takakura},\ and\
  \citenamefont {Ishimasa}}]{Deguchi2012}%
  \BibitemOpen
  \bibfield  {author} {\bibinfo {author} {\bibfnamefont {K.}~\bibnamefont
  {Deguchi}}, \bibinfo {author} {\bibfnamefont {S.}~\bibnamefont {Matsukawa}},
  \bibinfo {author} {\bibfnamefont {N.~K.}\ \bibnamefont {Sato}}, \bibinfo
  {author} {\bibfnamefont {T.}~\bibnamefont {Hattori}}, \bibinfo {author}
  {\bibfnamefont {K.}~\bibnamefont {Ishida}}, \bibinfo {author} {\bibfnamefont
  {H.}~\bibnamefont {Takakura}}, \ and\ \bibinfo {author} {\bibfnamefont
  {T.}~\bibnamefont {Ishimasa}},\ }\href {https://doi.org/10.1038/nmat3432}
  {\bibfield  {journal} {\bibinfo  {journal} {Nature Materials}\ }\textbf
  {\bibinfo {volume} {11}},\ \bibinfo {pages} {1013 EP } (\bibinfo {year}
  {2012})}\BibitemShut {NoStop}%
\bibitem [{\citenamefont {Hartman}\ \emph {et~al.}(2016)\citenamefont
  {Hartman}, \citenamefont {Chiu},\ and\ \citenamefont
  {Scalettar}}]{Scalettar16}%
  \BibitemOpen
  \bibfield  {author} {\bibinfo {author} {\bibfnamefont {N.}~\bibnamefont
  {Hartman}}, \bibinfo {author} {\bibfnamefont {W.-T.}\ \bibnamefont {Chiu}}, \
  and\ \bibinfo {author} {\bibfnamefont {R.~T.}\ \bibnamefont {Scalettar}},\
  }\href {\doibase 10.1103/PhysRevB.93.235143} {\bibfield  {journal} {\bibinfo
  {journal} {Phys. Rev. B}\ }\textbf {\bibinfo {volume} {93}},\ \bibinfo
  {pages} {235143} (\bibinfo {year} {2016})}\BibitemShut {NoStop}%
\bibitem [{\citenamefont {Collins}\ \emph {et~al.}(2017)\citenamefont
  {Collins}, \citenamefont {Witte}, \citenamefont {Silverman}, \citenamefont
  {Green},\ and\ \citenamefont {Gomes}}]{Collins2017}%
  \BibitemOpen
  \bibfield  {author} {\bibinfo {author} {\bibfnamefont {L.~C.}\ \bibnamefont
  {Collins}}, \bibinfo {author} {\bibfnamefont {T.~G.}\ \bibnamefont {Witte}},
  \bibinfo {author} {\bibfnamefont {R.}~\bibnamefont {Silverman}}, \bibinfo
  {author} {\bibfnamefont {D.~B.}\ \bibnamefont {Green}}, \ and\ \bibinfo
  {author} {\bibfnamefont {K.~K.}\ \bibnamefont {Gomes}},\ }\href {\doibase
  10.1038/ncomms15961} {\bibfield  {journal} {\bibinfo  {journal} {Nature
  Communications}\ }\textbf {\bibinfo {volume} {8}},\ \bibinfo {pages} {15961}
  (\bibinfo {year} {2017})}\BibitemShut {NoStop}%
\bibitem [{\citenamefont {Lenz}(2002)}]{Lenz-zero-weight}%
  \BibitemOpen
  \bibfield  {author} {\bibinfo {author} {\bibfnamefont {D.}~\bibnamefont
  {Lenz}},\ }\href {\doibase 10.1007/s002200200624} {\bibfield  {journal}
  {\bibinfo  {journal} {Communications in Mathematical Physics}\ }\textbf
  {\bibinfo {volume} {227}},\ \bibinfo {pages} {119} (\bibinfo {year}
  {2002})}\BibitemShut {NoStop}%
\bibitem [{\citenamefont {Damanik}\ and\ \citenamefont
  {Lenz}(1999)}]{Damanik1999}%
  \BibitemOpen
  \bibfield  {author} {\bibinfo {author} {\bibfnamefont {D.}~\bibnamefont
  {Damanik}}\ and\ \bibinfo {author} {\bibfnamefont {D.}~\bibnamefont {Lenz}},\
  }\href {\doibase 10.1007/s002200050742} {\bibfield  {journal} {\bibinfo
  {journal} {Communications in Mathematical Physics}\ }\textbf {\bibinfo
  {volume} {207}},\ \bibinfo {pages} {687} (\bibinfo {year}
  {1999})}\BibitemShut {NoStop}%
\bibitem [{\citenamefont {Mandel}\ and\ \citenamefont
  {Lifshitz}(2008)}]{square-and-cubic-fib}%
  \BibitemOpen
  \bibfield  {author} {\bibinfo {author} {\bibfnamefont {S.~E.-D.}\
  \bibnamefont {Mandel}}\ and\ \bibinfo {author} {\bibfnamefont
  {R.}~\bibnamefont {Lifshitz}},\ }\href {\doibase 10.1080/14786430802070805}
  {\bibfield  {journal} {\bibinfo  {journal} {Philosophical Magazine}\ }\textbf
  {\bibinfo {volume} {88}},\ \bibinfo {pages} {2261} (\bibinfo {year}
  {2008})}\BibitemShut {NoStop}%
\bibitem [{\citenamefont {Sire}(1989)}]{octagonal}%
  \BibitemOpen
  \bibfield  {author} {\bibinfo {author} {\bibfnamefont {C.}~\bibnamefont
  {Sire}},\ }\href {http://stacks.iop.org/0295-5075/10/i=5/a=016} {\bibfield
  {journal} {\bibinfo  {journal} {EPL (Europhysics Letters)}\ }\textbf
  {\bibinfo {volume} {10}},\ \bibinfo {pages} {483} (\bibinfo {year}
  {1989})}\BibitemShut {NoStop}%
\bibitem [{\citenamefont {Fujiwara}\ and\ \citenamefont
  {Yokokawa}(1991)}]{PhysRevLett.66.333}%
  \BibitemOpen
  \bibfield  {author} {\bibinfo {author} {\bibfnamefont {T.}~\bibnamefont
  {Fujiwara}}\ and\ \bibinfo {author} {\bibfnamefont {T.}~\bibnamefont
  {Yokokawa}},\ }\href {\doibase 10.1103/PhysRevLett.66.333} {\bibfield
  {journal} {\bibinfo  {journal} {Phys. Rev. Lett.}\ }\textbf {\bibinfo
  {volume} {66}},\ \bibinfo {pages} {333} (\bibinfo {year} {1991})}\BibitemShut
  {NoStop}%
\bibitem [{\citenamefont {Tang}\ \emph {et~al.}(1997)\citenamefont {Tang},
  \citenamefont {Hill}, \citenamefont {Wonnell}, \citenamefont {Poon},\ and\
  \citenamefont {Wu}}]{PhysRevLett.79.1070}%
  \BibitemOpen
  \bibfield  {author} {\bibinfo {author} {\bibfnamefont {X.-P.}\ \bibnamefont
  {Tang}}, \bibinfo {author} {\bibfnamefont {E.~A.}\ \bibnamefont {Hill}},
  \bibinfo {author} {\bibfnamefont {S.~K.}\ \bibnamefont {Wonnell}}, \bibinfo
  {author} {\bibfnamefont {S.~J.}\ \bibnamefont {Poon}}, \ and\ \bibinfo
  {author} {\bibfnamefont {Y.}~\bibnamefont {Wu}},\ }\href {\doibase
  10.1103/PhysRevLett.79.1070} {\bibfield  {journal} {\bibinfo  {journal}
  {Phys. Rev. Lett.}\ }\textbf {\bibinfo {volume} {79}},\ \bibinfo {pages}
  {1070} (\bibinfo {year} {1997})}\BibitemShut {NoStop}%
\bibitem [{\citenamefont {Kirihara}\ \emph {et~al.}(2003)\citenamefont
  {Kirihara}, \citenamefont {Nagata}, \citenamefont {Kimura}, \citenamefont
  {Kato}, \citenamefont {Takata}, \citenamefont {Nishibori},\ and\
  \citenamefont {Sakata}}]{PhysRevB.68.014205}%
  \BibitemOpen
  \bibfield  {author} {\bibinfo {author} {\bibfnamefont {K.}~\bibnamefont
  {Kirihara}}, \bibinfo {author} {\bibfnamefont {T.}~\bibnamefont {Nagata}},
  \bibinfo {author} {\bibfnamefont {K.}~\bibnamefont {Kimura}}, \bibinfo
  {author} {\bibfnamefont {K.}~\bibnamefont {Kato}}, \bibinfo {author}
  {\bibfnamefont {M.}~\bibnamefont {Takata}}, \bibinfo {author} {\bibfnamefont
  {E.}~\bibnamefont {Nishibori}}, \ and\ \bibinfo {author} {\bibfnamefont
  {M.}~\bibnamefont {Sakata}},\ }\href {\doibase 10.1103/PhysRevB.68.014205}
  {\bibfield  {journal} {\bibinfo  {journal} {Phys. Rev. B}\ }\textbf {\bibinfo
  {volume} {68}},\ \bibinfo {pages} {014205} (\bibinfo {year}
  {2003})}\BibitemShut {NoStop}%
\bibitem [{\citenamefont {Kohmoto}\ and\ \citenamefont
  {Sutherland}(1986)}]{Kohmoto_Sutherland}%
  \BibitemOpen
  \bibfield  {author} {\bibinfo {author} {\bibfnamefont {M.}~\bibnamefont
  {Kohmoto}}\ and\ \bibinfo {author} {\bibfnamefont {B.}~\bibnamefont
  {Sutherland}},\ }\href {\doibase 10.1103/PhysRevLett.56.2740} {\bibfield
  {journal} {\bibinfo  {journal} {Phys. Rev. Lett.}\ }\textbf {\bibinfo
  {volume} {56}},\ \bibinfo {pages} {2740} (\bibinfo {year}
  {1986})}\BibitemShut {NoStop}%
\bibitem [{\citenamefont {Arai}\ \emph {et~al.}(1988)\citenamefont {Arai},
  \citenamefont {Tokihiro}, \citenamefont {Fujiwara},\ and\ \citenamefont
  {Kohmoto}}]{nine_percent}%
  \BibitemOpen
  \bibfield  {author} {\bibinfo {author} {\bibfnamefont {M.}~\bibnamefont
  {Arai}}, \bibinfo {author} {\bibfnamefont {T.}~\bibnamefont {Tokihiro}},
  \bibinfo {author} {\bibfnamefont {T.}~\bibnamefont {Fujiwara}}, \ and\
  \bibinfo {author} {\bibfnamefont {M.}~\bibnamefont {Kohmoto}},\ }\href
  {\doibase 10.1103/PhysRevB.38.1621} {\bibfield  {journal} {\bibinfo
  {journal} {Phys. Rev. B}\ }\textbf {\bibinfo {volume} {38}},\ \bibinfo
  {pages} {1621} (\bibinfo {year} {1988})}\BibitemShut {NoStop}%
\bibitem [{\citenamefont {Koga}\ and\ \citenamefont
  {Tsunetsugu}(2017)}]{exact-number}%
  \BibitemOpen
  \bibfield  {author} {\bibinfo {author} {\bibfnamefont {A.}~\bibnamefont
  {Koga}}\ and\ \bibinfo {author} {\bibfnamefont {H.}~\bibnamefont
  {Tsunetsugu}},\ }\href {\doibase 10.1103/PhysRevB.96.214402} {\bibfield
  {journal} {\bibinfo  {journal} {Phys. Rev. B}\ }\textbf {\bibinfo {volume}
  {96}},\ \bibinfo {pages} {214402} (\bibinfo {year} {2017})}\BibitemShut
  {NoStop}%
\bibitem [{\citenamefont {Flicker}\ \emph {et~al.}(2020)\citenamefont
  {Flicker}, \citenamefont {Simon},\ and\ \citenamefont
  {Parameswaran}}]{dimers}%
  \BibitemOpen
  \bibfield  {author} {\bibinfo {author} {\bibfnamefont {F.}~\bibnamefont
  {Flicker}}, \bibinfo {author} {\bibfnamefont {S.~H.}\ \bibnamefont {Simon}},
  \ and\ \bibinfo {author} {\bibfnamefont {S.~A.}\ \bibnamefont
  {Parameswaran}},\ }\href {\doibase 10.1103/PhysRevX.10.011005} {\bibfield
  {journal} {\bibinfo  {journal} {Phys. Rev. X}\ }\textbf {\bibinfo {volume}
  {10}},\ \bibinfo {pages} {011005} (\bibinfo {year} {2020})}\BibitemShut
  {NoStop}%
\bibitem [{\citenamefont {Hatakeyama}\ and\ \citenamefont
  {Kamimura}(1987)}]{HATAKEYAMA198779}%
  \BibitemOpen
  \bibfield  {author} {\bibinfo {author} {\bibfnamefont {T.}~\bibnamefont
  {Hatakeyama}}\ and\ \bibinfo {author} {\bibfnamefont {H.}~\bibnamefont
  {Kamimura}},\ }\href {\doibase https://doi.org/10.1016/0038-1098(87)91116-1}
  {\bibfield  {journal} {\bibinfo  {journal} {Solid State Communications}\
  }\textbf {\bibinfo {volume} {62}},\ \bibinfo {pages} {79 } (\bibinfo {year}
  {1987})}\BibitemShut {NoStop}%
\bibitem [{\citenamefont {Aoyama}\ and\ \citenamefont
  {Odagaki}(1987)}]{aoyama1987eight}%
  \BibitemOpen
  \bibfield  {author} {\bibinfo {author} {\bibfnamefont {H.}~\bibnamefont
  {Aoyama}}\ and\ \bibinfo {author} {\bibfnamefont {T.}~\bibnamefont
  {Odagaki}},\ }\href@noop {} {\bibfield  {journal} {\bibinfo  {journal}
  {Journal of statistical physics}\ }\textbf {\bibinfo {volume} {48}},\
  \bibinfo {pages} {503} (\bibinfo {year} {1987})}\BibitemShut {NoStop}%
\bibitem [{\citenamefont {You}\ \emph {et~al.}(1992)\citenamefont {You},
  \citenamefont {Yan}, \citenamefont {Zhong},\ and\ \citenamefont
  {Yan}}]{You_1992}%
  \BibitemOpen
  \bibfield  {author} {\bibinfo {author} {\bibfnamefont {J.~Q.}\ \bibnamefont
  {You}}, \bibinfo {author} {\bibfnamefont {J.~R.}\ \bibnamefont {Yan}},
  \bibinfo {author} {\bibfnamefont {J.~X.}\ \bibnamefont {Zhong}}, \ and\
  \bibinfo {author} {\bibfnamefont {X.~H.}\ \bibnamefont {Yan}},\ }\href
  {\doibase 10.1209/0295-5075/17/3/008} {\bibfield  {journal} {\bibinfo
  {journal} {Europhysics Letters ({EPL})}\ }\textbf {\bibinfo {volume} {17}},\
  \bibinfo {pages} {231} (\bibinfo {year} {1992})}\BibitemShut {NoStop}%
\bibitem [{\citenamefont {Sutherland}(1986)}]{PhysRevB.34.3904}%
  \BibitemOpen
  \bibfield  {author} {\bibinfo {author} {\bibfnamefont {B.}~\bibnamefont
  {Sutherland}},\ }\href {\doibase 10.1103/PhysRevB.34.3904} {\bibfield
  {journal} {\bibinfo  {journal} {Phys. Rev. B}\ }\textbf {\bibinfo {volume}
  {34}},\ \bibinfo {pages} {3904} (\bibinfo {year} {1986})}\BibitemShut
  {NoStop}%
\bibitem [{\citenamefont {De~Bruijn}(1981)}]{de1981ned}%
  \BibitemOpen
  \bibfield  {author} {\bibinfo {author} {\bibfnamefont {N.}~\bibnamefont
  {De~Bruijn}},\ }in\ \href@noop {} {\emph {\bibinfo {booktitle} {Proc. Ser.
  A}}},\ Vol.~\bibinfo {volume} {43}\ (\bibinfo {year} {1981})\ p.~\bibinfo
  {pages} {53}\BibitemShut {NoStop}%
\bibitem [{\citenamefont {Kumar}\ \emph {et~al.}(1986)\citenamefont {Kumar},
  \citenamefont {Sahoo},\ and\ \citenamefont {Athithan}}]{site_fractions}%
  \BibitemOpen
  \bibfield  {author} {\bibinfo {author} {\bibfnamefont {V.}~\bibnamefont
  {Kumar}}, \bibinfo {author} {\bibfnamefont {D.}~\bibnamefont {Sahoo}}, \ and\
  \bibinfo {author} {\bibfnamefont {G.}~\bibnamefont {Athithan}},\ }\href
  {\doibase 10.1103/PhysRevB.34.6924} {\bibfield  {journal} {\bibinfo
  {journal} {Phys. Rev. B}\ }\textbf {\bibinfo {volume} {34}},\ \bibinfo
  {pages} {6924} (\bibinfo {year} {1986})}\BibitemShut {NoStop}%
\bibitem [{\citenamefont {Chiu}\ \emph {et~al.}(2016)\citenamefont {Chiu},
  \citenamefont {Teo}, \citenamefont {Schnyder},\ and\ \citenamefont
  {Ryu}}]{Ryu2016}%
  \BibitemOpen
  \bibfield  {author} {\bibinfo {author} {\bibfnamefont {C.-K.}\ \bibnamefont
  {Chiu}}, \bibinfo {author} {\bibfnamefont {J.~C.~Y.}\ \bibnamefont {Teo}},
  \bibinfo {author} {\bibfnamefont {A.~P.}\ \bibnamefont {Schnyder}}, \ and\
  \bibinfo {author} {\bibfnamefont {S.}~\bibnamefont {Ryu}},\ }\href {\doibase
  10.1103/RevModPhys.88.035005} {\bibfield  {journal} {\bibinfo  {journal}
  {Rev. Mod. Phys.}\ }\textbf {\bibinfo {volume} {88}},\ \bibinfo {pages}
  {035005} (\bibinfo {year} {2016})}\BibitemShut {NoStop}%
\bibitem [{\citenamefont {de~Laissardi{\`{e}}re}\ \emph
  {et~al.}(2014)\citenamefont {de~Laissardi{\`{e}}re}, \citenamefont
  {Sz{\'{a}}ll{\'{a}}s},\ and\ \citenamefont
  {Mayou}}]{TramblydeLaissardire2014}%
  \BibitemOpen
  \bibfield  {author} {\bibinfo {author} {\bibfnamefont {G.~T.}\ \bibnamefont
  {de~Laissardi{\`{e}}re}}, \bibinfo {author} {\bibfnamefont {A.}~\bibnamefont
  {Sz{\'{a}}ll{\'{a}}s}}, \ and\ \bibinfo {author} {\bibfnamefont
  {D.}~\bibnamefont {Mayou}},\ }\href {\doibase 10.12693/aphyspola.126.617}
  {\bibfield  {journal} {\bibinfo  {journal} {Acta Physica Polonica A}\
  }\textbf {\bibinfo {volume} {126}},\ \bibinfo {pages} {617} (\bibinfo {year}
  {2014})}\BibitemShut {NoStop}%
\bibitem [{\citenamefont {Levine}\ and\ \citenamefont
  {Steinhardt}(1984)}]{original-theory}%
  \BibitemOpen
  \bibfield  {author} {\bibinfo {author} {\bibfnamefont {D.}~\bibnamefont
  {Levine}}\ and\ \bibinfo {author} {\bibfnamefont {P.~J.}\ \bibnamefont
  {Steinhardt}},\ }\href {\doibase 10.1103/PhysRevLett.53.2477} {\bibfield
  {journal} {\bibinfo  {journal} {Phys. Rev. Lett.}\ }\textbf {\bibinfo
  {volume} {53}},\ \bibinfo {pages} {2477} (\bibinfo {year}
  {1984})}\BibitemShut {NoStop}%
\bibitem [{\citenamefont {{Sire, Cl\'ement}}\ and\ \citenamefont {{Mosseri,
  R\'emy}}(1989)}]{gap-labelling}%
  \BibitemOpen
  \bibfield  {author} {\bibinfo {author} {\bibnamefont {{Sire, Cl\'ement}}}\
  and\ \bibinfo {author} {\bibnamefont {{Mosseri, R\'emy}}},\ }\href {\doibase
  10.1051/jphys:0198900500240344700} {\bibfield  {journal} {\bibinfo  {journal}
  {J. Phys. France}\ }\textbf {\bibinfo {volume} {50}},\ \bibinfo {pages}
  {3447} (\bibinfo {year} {1989})}\BibitemShut {NoStop}%
\bibitem [{\citenamefont {Merlin}\ \emph {et~al.}(1985)\citenamefont {Merlin},
  \citenamefont {Bajema}, \citenamefont {Clarke}, \citenamefont {Juang},\ and\
  \citenamefont {Bhattacharya}}]{PhysRevLett.55.1768}%
  \BibitemOpen
  \bibfield  {author} {\bibinfo {author} {\bibfnamefont {R.}~\bibnamefont
  {Merlin}}, \bibinfo {author} {\bibfnamefont {K.}~\bibnamefont {Bajema}},
  \bibinfo {author} {\bibfnamefont {R.}~\bibnamefont {Clarke}}, \bibinfo
  {author} {\bibfnamefont {F.~Y.}\ \bibnamefont {Juang}}, \ and\ \bibinfo
  {author} {\bibfnamefont {P.~K.}\ \bibnamefont {Bhattacharya}},\ }\href
  {\doibase 10.1103/PhysRevLett.55.1768} {\bibfield  {journal} {\bibinfo
  {journal} {Phys. Rev. Lett.}\ }\textbf {\bibinfo {volume} {55}},\ \bibinfo
  {pages} {1768} (\bibinfo {year} {1985})}\BibitemShut {NoStop}%
\end{thebibliography}%

\end{document}